\documentclass{article}
\usepackage{chemformula} 
\usepackage[T1]{fontenc} 
\usepackage{lmodern}			

\usepackage{graphicx}
\usepackage{rotating}
\usepackage{tikz}

\usepackage[utf8]{inputenc}		
\usepackage{color}				
\usepackage{graphics,graphicx}			
\usepackage{microtype} 			
\usepackage{relsize}				
\usepackage{subfigure}
\usepackage{caption}
\usepackage{subcaption}
\usepackage[abs]{overpic}
\usepackage{hyperref}
\usepackage{hypcap}
\usepackage{amsmath,amssymb,amsthm}	
\usepackage{pdflscape} 		
\usepackage{multirow}			
\usepackage{adjustbox}			
\usepackage{mathtools}
\usepackage{lipsum}	 
\usepackage{cite}
\usepackage{float}
\usepackage{multicol}

\providecommand{\keywords}[1]
{
  \small	
  \textbf{water, lattice, confined} #1
}

\title{ Wetting in Associating Lattice Gas Model Confined by Hydrophilic Walls }

\author{Tássylla O. Fonseca$^{1*}$, Bruno H. S. Mendonça$^{1*}$, Elizane E. de Moraes$^{2}$, \\Alan B. de Oliveira$^{3}$ and Marcia C. Barbosa \\
        \\ 
        \hspace{-0.5cm}
        \scriptsize{\textit{$^{1}$ Departamento de Física, Universidade Federal de Minas Gerais, 31270-901, Belo Horizonte, MG, Brasil}} \\
        \hspace{-0.5cm}
        \scriptsize{\textit{$^{2}$ Instituto de Física, Universidade Federal da Bahia, 40170-115, Salvador, BA, Brasil}} \\
        \hspace{-0.5cm}
        \scriptsize{\textit{$^{3}$ Departamento de Física, Universidade Federal de Ouro Preto, 35400-000, Ouro Preto, MG, Brasil}} \\
        \hspace{-0.5cm}
        \scriptsize{\textit{ Instituto de Física, Universidade Federal do Rio Grande do Sul, 91501-970, Porto Alegre, RS, Brasil}} \\
        }

\begin{document}

\maketitle

\begin{abstract}
Through Monte Carlo simulations and the Associating Lattice Gas Model, the phases of a two-dimensional fluid under 
hydrophilic confinement are evaluated. The model, in its unconfined version, reproduces the anomalous behavior of water 
regarding its density, diffusion, and solubility, among other dynamic and thermodynamic properties. Extreme confinements 
suppress phase transitions since fluctuations suppress ordering. The fluid under hydrophilic confinement forms a single 
wetting layer that gradually wets the wall. From the wetting layer, the low-density liquid structure is formed. 
The confined fluid presents a first-order liquid-liquid transition, but always at lower temperatures than that observed in the bulk. \\
\end{abstract}

\keywords{}

\section{Introduction}

Nanoconfined water has applications in medicine, such as controlled drug release~\cite{barroug2004interactions,pai2006pharmaceutical}, genetic engineering~\cite{pai2006pharmaceutical}, 
construction of new and more effective vaccines~\cite{pantarotto2003immunization}, cancer treatment~\cite{kam2005carbon}, artificial implants~\cite{ding2001recent}, and sustainable 
development, such as filters for water treatment~\cite{wang2010natural,babel2003low}, application in the manufacture of second-generation ethanol, 
or cellulosic ethanol~\cite{fonseca2017freezing,Driemeier2012}, phase separation~\cite{Gelb2006}, manufacture of nanomaterials~\cite{Deville2006} and also the search for the 
theoretical second critical point supposedly present in the supercooled region of the water phase diagram~\cite{ZheWang2015}.

Even with a vast literature, with numerous studies on this topic, there are still many misunderstood behaviors. 
Therefore, continued research into water confined in nanomaterials becomes essential.

For different types of confinement studied, experiments in cylindrical pores~\cite{Erko2011,Morishige1999}, simulations in porous matrices~\cite{Gallo2007,Pizio2009}, 
simulations in carbon nanotubes~\cite{Koga2001,Hummer2001}, experiments~\cite{Bellissent1996,Zanotti2005} and simulations on rough surfaces~\cite{Koga2005,Choudhury2010} and simulations 
on flat plates \cite{Meyer1999,Kumar2005}, differences observed between the behavior of bulk (unconfined) water and that of confined water. 
For subnanometric systems, water is more mobile and presents a flow that violates the laws of classical hydrodynamics.

In addition to the discrepancies when considering bulk water and water under confinement, distinct behaviors are expected depending 
on the degree of confinement and the nature of the confining material, hydrophilic or hydrophobic. The formation of water layers has 
been reported for several different types of confinement~\cite{Zangi2003,Gallo2010}, in some cases transitions linked to the change in the number of 
layers are observed when the degree of confinement of the system varies \cite{Zangi2003,Giovambattista2009} and structured phases in the contact layer \cite{Nanok2009,Krott2014,Bordin2014}. 
Cylindrical silica nanopores lower the equilibrium melting temperature of water, pores with a radius greater than 50 nm have essentially 
the same melting temperature as bulk water, while pores with a radius of 1 nm, or even smaller, do not exhibit a first-order melting transition~\cite{Findenegg2008,Allenhnert2008}. 
Experiments on silica nanotubes show that the water occupying the innermost region of the pore, also called free water, freezes at a lower 
temperature than the bulk water. In contrast, the water in contact with the confining wall, or bound water, remains liquid \cite{Giovambattista2006,Molinero2012,Morishige1999,Stapf1995}. 
Some studies suggest that free water, when it freezes, forms metastable cubic ice instead of the hexagonal ice observed in bulk water~\cite{STEYTLER1983, Liu2006, OVERLOOP1993, Morishige1999,Morishige2003}, 
but Baker et. al. showed that the type of ice formed inside the nanotube depends on the conditions and the nature of the confining material~\cite{Baker1997}.

Other studies have shown that the melting temperatures of nanoconfined ice are sensitive to pore diameter but are 
not affected by pore surface functionalization. Findenegg et. al. used calorimetry to study the melting ice in nanopores 
of SBA-15 silica functionalized with carboxylic acid, sulfonic acid, or phosphonic acid. They found that none of the surface 
modifiers changed the melting temperature of confined ice. Deschamps et al.~\cite{Deschamps2010} and Jelassi et al.~\cite{Jelassi2010} studied the melting 
of ice in MCM-41 silica pores and in mesoporous silica gel, in its hydrophilic form and functionalized with hydrophobic groups. 
They found very little difference in the melting temperature of nanoconfined ice in hydrophilic and hydrophobic pores.

Simulations also show contradictory results regarding the effects of the type of water-wall interaction on the transition 
temperature of confined water. While the results of the SPC/E model show that this temperature between hydrophobic plates 
is lower than in the bulk and higher than for hydrophilic confinement~\cite{Giovambattista2009}, for the mW model, no difference is found~\cite{Molinero2012}.

Therefore, confined water at the nanoscale presents a series of behaviors that are not observed in unconfined water, 
and these behaviors can vary depending on the confining material, type of confinement (hydrophobic or hydrophilic walls) 
and degree of confinement (size of the confining matrix). Under confinement, in addition to the water-water interaction, 
the water-wall interaction must be accounted for. Since the behavior of water changes at the interface, 
it cannot be treated in the same way as bulk water. Therefore, knowing the confining material, its structure, 
and how it interacts with the fluid is essential for study of confined systems~\cite{Gavazzoni2017}.

In particular, since the experimental detection of the two liquid phases of water~\cite{ZheWang2015} is a difficult task, 
since the system crystallizes, and since confinement reduces the crystallization temperature, experiments have 
recently been carried out to observe the two phases in a confined system~\cite{Chen2004,Xu2005}. In these experiments, a dynamic 
transition was observed that signals the existence of the two liquid phases. This observation that associates dynamic 
transition with criticality and the two phases was also corroborated by the minimal associative lattice gas model~\cite{Szortyka2010134904,Szortyka2009}.

However, applying testing hypotheses for unconfined systems from confined systems must be cautious. 
Can the physics be generalized to both systems? What would be the confinement limit at which the physical behaviors would converge? 
How does the transition temperature of confined water depend on the nature of the confining wall and the size of the confinement, 
consequently, how is the phase diagram of water in the chemical potential vs. temperature plane affected in comparison to 
the unconfined system? Can we quantify structural frustration? The answers to these questions are still under debate. 
They are not simple answers, but they are fundamental for describing the properties that are specific to confinement 
and the properties with universal characteristics.

To contribute to the explanation of these phenomena, we propose a two-dimensional Associative Lattice Gas Model for hydrophilic 
confinement, where the degree of confinement and the length of the confining wall lines vary. The choice of a simple model aims 
to understand whether confinement leads to the emergence of new phases and whether the type of wall would influence this process. 
The idea is to understand specific properties of confinement, as well as to identify at which confinement limits we can use the 
confined system as an adequate representation of the unconfined system, and thus enable its use to investigate the two liquid phases 
and the potential criticality present in them.

\section{Model and Methods}\label{sec:2}

We investigated a fluid under two-dimensional hydrophilic confinement through the 2D model Associative Lattice Gas, proposed by 
Henriques and Barbosa~\cite{Henriques2005,Szortyka2009} for bulk water. And later used for water confined by hydrophobic walls~\cite{Fonseca2019}.
We tested whether the confinement system presents new phases in addition to the displacement of the bulk properties.
\begin{figure}[H]
\begin{center}
\includegraphics[scale=1.95]{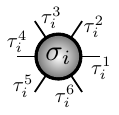}
\caption{ Schematic representation of a particle of the lattice with its occupational, 
$\sigma_i$, and orientational variables, ${\tau_{i}}^A$, where its six arms are presented, 
$A$= 1, 2, 3... 6. Reproduced from reference~\cite{Fonseca2019}. }
\label{fig:1}
\end{center}
\end{figure}

\begin{figure}[H]
\begin{center}
\includegraphics[scale=1.0]{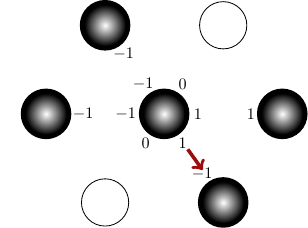}
\caption{ Representation of the central particle with its six arms and its first six neighbors with arms pointing to it. 
Note the hydrogen bond, represented by the arrow, coming out of the donor arm of the central particle and pointing to 
the acceptor arm of its neighbor (arms with complementary orientations, $\tau_{i}^{A}$$\tau_{i}^ {B}$= -1). }
\label{fig:2}
\end{center}
\end{figure}

\begin{figure}[H]
\begin{center}
\includegraphics[scale=1.0]{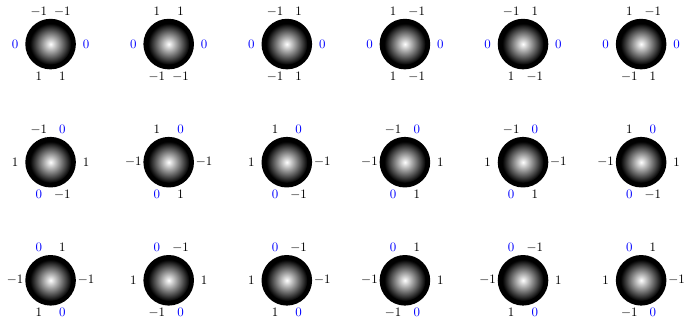}
\caption{Directions of the inert arms (these are diagonally opposite).}
\label{fig:3}
\end{center}
\end{figure}

A triangular lattice of dimension $L_xL_y$ describes the fluid where $L_y$ is under periodic boundary conditions, 
while $L_x$ is finite to reproduce confinement. Each site of the lattice can accommodate a particle and, for this purpose, 
is associated with two variables: an occupation variable linking the presence or absence of particles to the site 
(with $\sigma_i$= 1 or 0, in the case of particles fluids, and $\xi_k $= 1 or 0 , for wall particles), and an arm variable, 
which represents the possibility of hydrogen bond formation between two neighboring particles. Each fluid particle has six arms, 
${\tau_{i}}^A$, where $A$= 1, ..., 6 (where $i$ represents the particle index and $A$ the arm variable) to represent the 
orientational degrees of freedom present, for example, in water, as illustrated by Figure~\ref{fig:1}. In the case of water, 
${\tau_{i}}^A$= 1 represents the electron donating arm (charge distribution at the oxygen vicinity), ${\tau_{i}}^A$= -1 
represents the electron acceptor arm (charge distribution at the hydrogen vicinity) and ${\tau_{i}}^A$= 0 represents nonbinding directions. 
The tetrahedral structure of water is represented when each particle has two acceptors, two donors, and two opposite inert arms, as illustrated 
in figure~\ref{fig:2}. Therefore has eighteen configuration states for each particle as indicated in figure~\ref{fig:3}. 
A bond is formed when two neighboring sites have arms with complementary orientations, ${\tau_{i}}^A{\tau_{i}}^B$= -1, i.e., 
the product of the arms equals -1 if a donor arm points to an acceptor arm. In the case of water is a hydrogen bond.

\begin{figure}[H]
\begin{center}
\subfigure{%
\hspace{1.25cm}
\begin{overpic}[scale=0.35,unit=1mm]{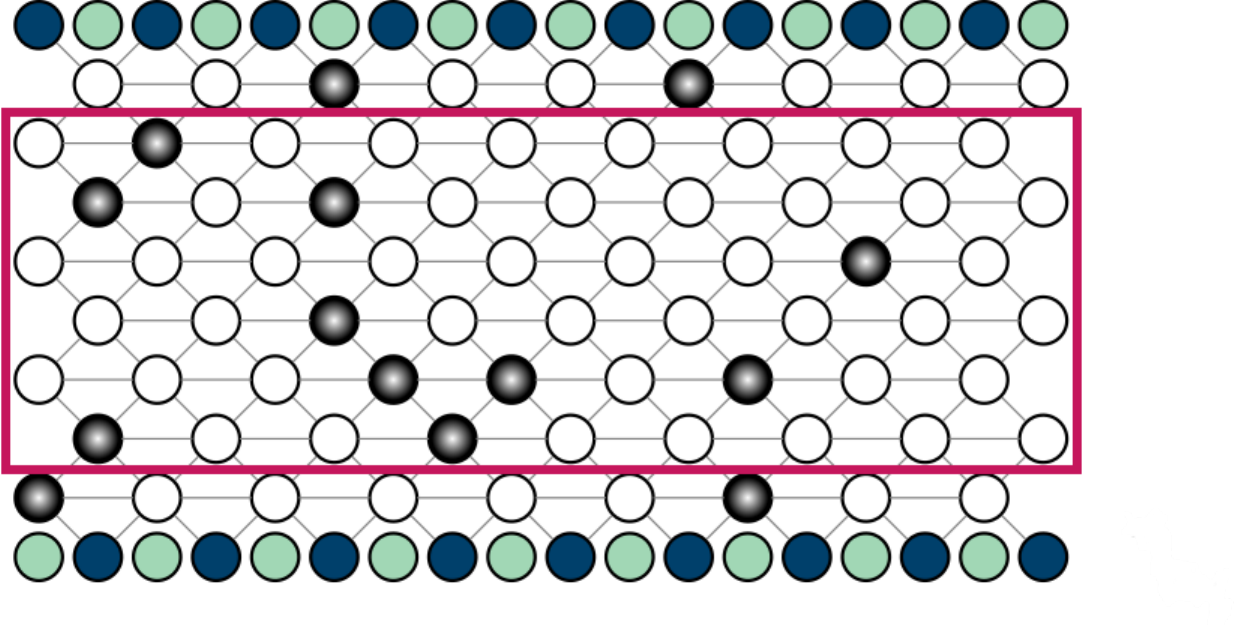}
\put(-3.5,37.5){{\parbox{0.4\linewidth}{
(a)
}}}
\end{overpic}}\vfill 
\subfigure{%
\hspace{1.35cm}
\begin{overpic}[scale=0.35,unit=1mm]{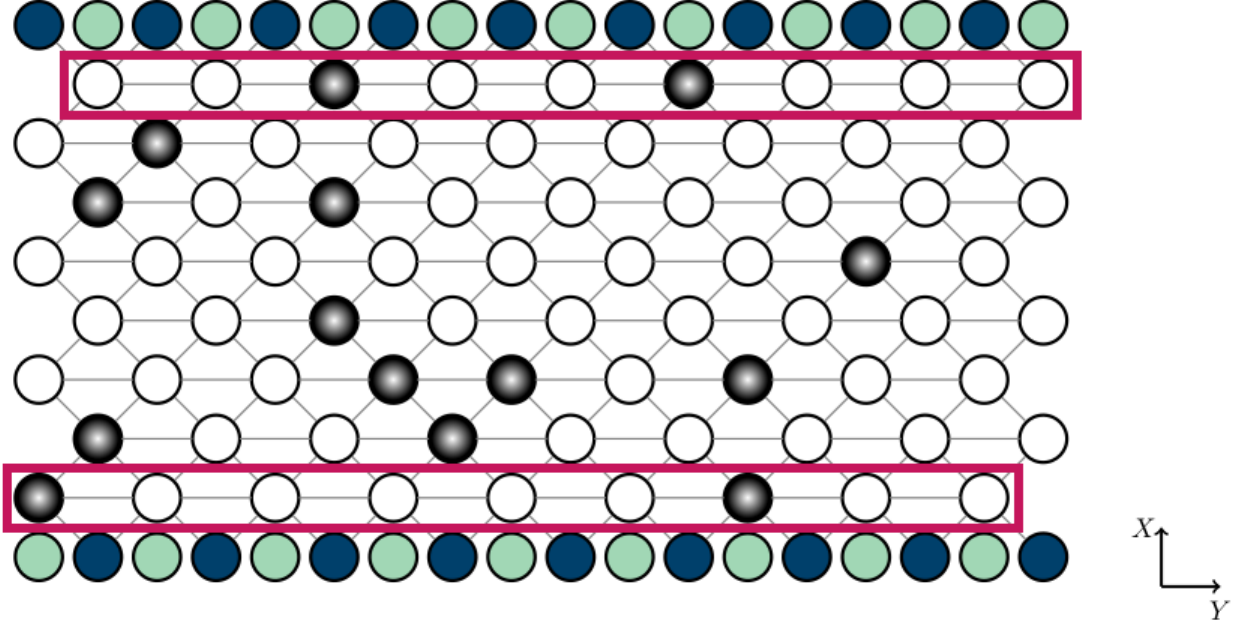}
\put(-3.5,37.5){{\parbox{0.4\linewidth}{
(b)
}}}
\end{overpic}}
\caption{ Schematic representation highlighting the two regions of the confined system. 
(a) Sites of the central layers and (b) sites belonging to the layers in contact with the 
confining walls. The light and dark green circles represent the wall, the filled black 
circles represent the water particles and the white circles represent the empty sites. Reproduced from reference~\cite{Fonseca2019}. }
\label{fig:4}
\end{center}
\end{figure}

The representation of fluid confined between two fixed lines of particles is depicted in Figure~\ref{fig:4}.
%

To simulate hydrophilic confinement, we investigated systems with confining wall particles, illustrated as light and 
dark green sites, which can exhibit arm interactions with the confined fluid. Wall particles with states ${\zeta_{k}}^B$=~-1 or~+1 are hydrophilic, 
representing hydrophilic confinements, for example, walls with polar groups. In this work, we investigated the 
hydrophilic confinement through the bond donor walls, i.e., ${\zeta_{k}}^B$=~+1.

The Hamiltonian for confinement consists of two contributions: the interaction between water molecules and the interaction 
between water molecules and wall particles. Therefore, we have divided the lattice into two parts: the sites belonging to the central 
layers and the sites in contact with the walls. The particles in the central layers are described by the Hamiltonian where the fluid-fluid 
interaction is over the first six neighbors, while the particles in the contact layer are described by the Hamiltonian where the fluid-fluid 
interaction is over four neighbors from two of the first six neighbors are wall particles~\cite{Fonseca2019}.

As there are variations in the number of particles in the system, is use the Grand Canonical ensemble. 
For the implementation of Monte Carlo, we, therefore, use the combination of Hamiltonians:
\begin{equation}\label{eq:ModeloConf-01}
{\hat{\mathcal{H}}}_{central}=(-v+2u){\sum_{\langle{i,k}\rangle}^{*}{\sigma_{i}\sigma_{k}}}+
u{\sum_{\langle{i,k}\rangle}^{*}{\sigma_{i}\sigma_{k}}}{\sum_{A=1}^{6}{\sum_{B^{*}=1}^{6}}{[(1-\tau_{i}^{A}\tau_{k}^{B})\tau_{i}^{A}\tau_{k}^{B}]}}
\end{equation}
\begin{equation}\label{eq:ModeloConf-02}
\begin{aligned}
{\hat{\mathcal{H}}}_{contact}=(-v+2u){\sum_{\langle{i,k}\rangle}^{4}{\sigma_{i}\sigma_{k}}}+
u{\sum_{\langle{i,k}\rangle}^{4}{\sigma_{i}\sigma_{k}}}{\sum_{A=1}^{4}{\sum_{B^{*}=1}^{4}}{[(1-\tau_{i}^{A}\tau_{k}^{B})\tau_{i}^{A}\tau_{k}^{B}]}}+\\
u_p{\sum_{\langle{i,k}\rangle}^{2}{\sigma_{i}\xi_{k}}}{\sum_{A=1}^{2}{\sum_{B^{*}=1}^{2}}{[(1-\tau_{i}^{A}\zeta_{k}^{B})\tau_{i}^{A}\zeta_{k}^{B}]}}
\end{aligned}
\end{equation}

In these descriptions of energy, $-v$ is the short-range energy between the first fluid-fluid neighbors, 
$2u$ describes the binding energy between two neighboring fluid particles, while $u_p$ is the fluid-wall interaction. 
Since $u$ and $v$ are positive, $u/v>$1/2, a repulsive interaction, $2u-v$, is generated when the first neighbors 
do not form a hydrogen bond, and the bonding neighbors, exhibit an attractive energy $ -v$. Here, $u/v$=~1, 
to understand the bonding behavior in liquids under confinement.
Fluid particles have $\sigma_i$= 0, 1 as occupancy variables and $\tau_i^A$ and $\tau_k^B$ equal to 0, $\pm$~1 representing their arm variables. 
While the occupancy variables of the wall particles are represented by $\xi_k$=~1, and the arm variables by $\zeta_k^B$. 
In the prior study, the focus was the hydrophobic confinement~\cite{Fonseca2019}, where the confining walls' particles 
cannot form hydrogen bonds with the fluid particles, leading to the arm variable of the wall particles assuming the 
value $\zeta_k^B$ = 0, as shown in Figure 5. In our current work, we analyze hydrophilic confinement, which entails 
the potential for hydrogen bonding between the fluid particles in the contact layers and the wall particles. 
This results in $\zeta_k^B \neq$~0, with values possibly equal to $\pm$~1. Specifically, we are exploring hydrophilic 
confinement through the bond donor walls, with the arm variables of the wall particles taking the value $\zeta_k^B$=~+1.
And, $u_p$= 2 is chosen to energetically favor the fluid-wall interaction over the fluid-fluid interaction.
\begin{figure}[H]
\begin{center}
\subfigure{%
\hspace{-0.85cm}
\begin{overpic}[scale=0.9,unit=1mm]{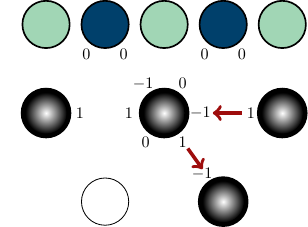}
\put(-3.5,37.5){{\parbox{0.4\linewidth}{
(a)
}}}
\end{overpic}}\vfill \vspace{0.7cm}
\subfigure{%
\hspace{-0.85cm}
\begin{overpic}[scale=0.9,unit=1mm]{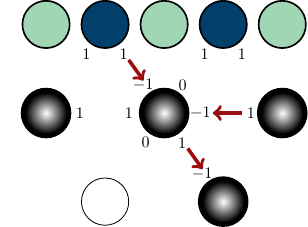}
\put(-3.5,37.5){{\parbox{0.4\linewidth}{
(b)
}}}
\end{overpic}}
\caption{ Representative scheme of the contact layer for the two types of confinement: (a) hydrophobic, the wall particles 
(circles in light blue and dark blue) cannot form hydrogen bonds with the confined fluid (for this, the two arms of the particles 
of walls are inert, i.e. $\zeta_k^B$=~0) and (b) hydrophilic, the wall particles are allowed to form hydrogen bonds with the fluid 
particles (the arms of the wall particles are electron donors, or i.e. $\zeta_k^B$=~+1). The black and white circles represent the 
confined fluid sites, which may or may not be occupied. The arrows, in red, represent hydrogen bonds. }
\label{fig:5}
\end{center}
\end{figure}

Employ reduced units for temperature and chemical potential,
\begin{equation}
\label{eq:04}
 \overline{T}= \frac{k_BT}{v}
\end{equation}
\begin{equation}
\label{eq:05}
 \overline{\mu}= \frac{\mu}{v}
\end{equation}

We performed Monte Carlo simulations using the Metropolis algorithm in the grand canonical ensemble.
At low chemical potential, the initial configuration employed was the empty lattice, while for high chemical potential stood lattice full. 
We analyzed differents initials configurations that were not difference in the final structure and density. 
Tested the confinement direction for different sizes, $L_y$= 10, 30, and 60, the value $L_y$= 30 is adopted 
why minimize the computational cost since the results for the density become independent of the system size relative to that direction. 
We investigated different degrees of confinement through $L_x$= 2, 4, 6, 8, 10, 12, and 14. Throughout the simulation, 
the particles belonging to the walls have fixed arm variable states.
Equilibration of the system involved $2$x$10^7$ Monte Carlo steps. The time required to de-correlate the measurements 
was $5$x$10^2$ Monte Carlo steps. Sampling consisted of $5$x$10^3$ measurements.

\vspace{0.5cm}
\begin{figure}[H]
\begin{center}
\subfigure{%
\hspace{1.25cm}
\begin{overpic}[scale=0.65,unit=1mm]{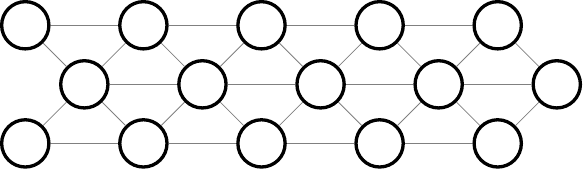}
\put(-3.5,20.5){{\parbox{0.4\linewidth}{
(a)
}}}
\end{overpic}}\vfill \vspace{0.7cm}
\subfigure{%
\hspace{1.35cm}
\begin{overpic}[scale=0.65,unit=1mm]{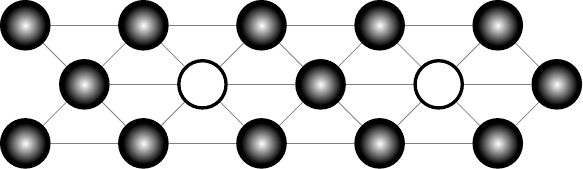}
\put(-3.5,20.5){{\parbox{0.4\linewidth}{
(b)
}}}
\end{overpic}}\vfill \vspace{0.7cm}
\subfigure{%
\hspace{1.35cm}
\begin{overpic}[scale=0.65,unit=1mm]{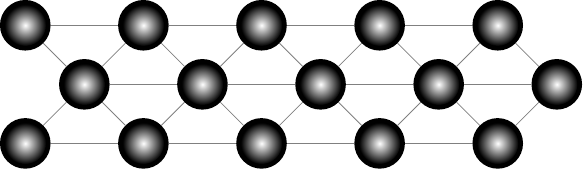}
\put(-3.5,20.5){{\parbox{0.4\linewidth}{
(c)
}}}
\end{overpic}}
\caption{ (a) In the gas phase, the whole lattice is empty. (b) In the low density phase, LDL, the lattice is 3/4 filled and particles are 
distributed over the lattice in such a way that the inert arms point only to the empty sites. There is no energy punishment, in this case. 
(c) In the high density phase, HDL, the lattice is full, and an energy punishment arises, because two inert arms point to filled sites. 
Reproduced from reference~\cite{Fonseca2019}. }
\label{fig:6}
\end{center}
\end{figure}
Henriques and Barbosa and later Szortyka et. al.~\cite{Henriques2005,Szortyka2009}, showed that the model 
presents distinct phases in the ground state: gaseous phase (GAS), at low chemical potentials, with $\rho$= 0 (empty lattice); 
the low-density liquid phase (LDL) where $3/4$ of the lattice is occupied $\rho$= 0.75, in intermediate chemical potentials; 
and the high-density liquid phase (HDL), at high chemical potentials, which is characterized by having 
all the lattice sites occupied by particles, that is, $\rho$= 1. There is coexistence between phases. 
At $\mu^*= -2v$, the GAS and LDL phases coexist with one energy per site $e_{gas-LDL}=E_{gas-LDL}/L^2=-3v/2$. 
And for higher chemical potential, $\mu^*= 6 + 8u/v$, the LDL phase coexists with the HDL phase with an energy per site 
of $E_{LDL-HDL}/L^2=e_{LDL-HDL}=-3v+2u$. In the LDL phase, all particles manage to form four hydrogen bonds, as there are 
neighboring sites that are empty, while in the HDL phase, with the entire lattice occupied, each particle can't form 
four hydrogen bonds with all its neighbors, so the system is penalized for having neighbors that do not
a bonding (shown in Figure~\ref{fig:6}).

\begin{figure}[H]
\begin{center}
\includegraphics[scale=0.3]{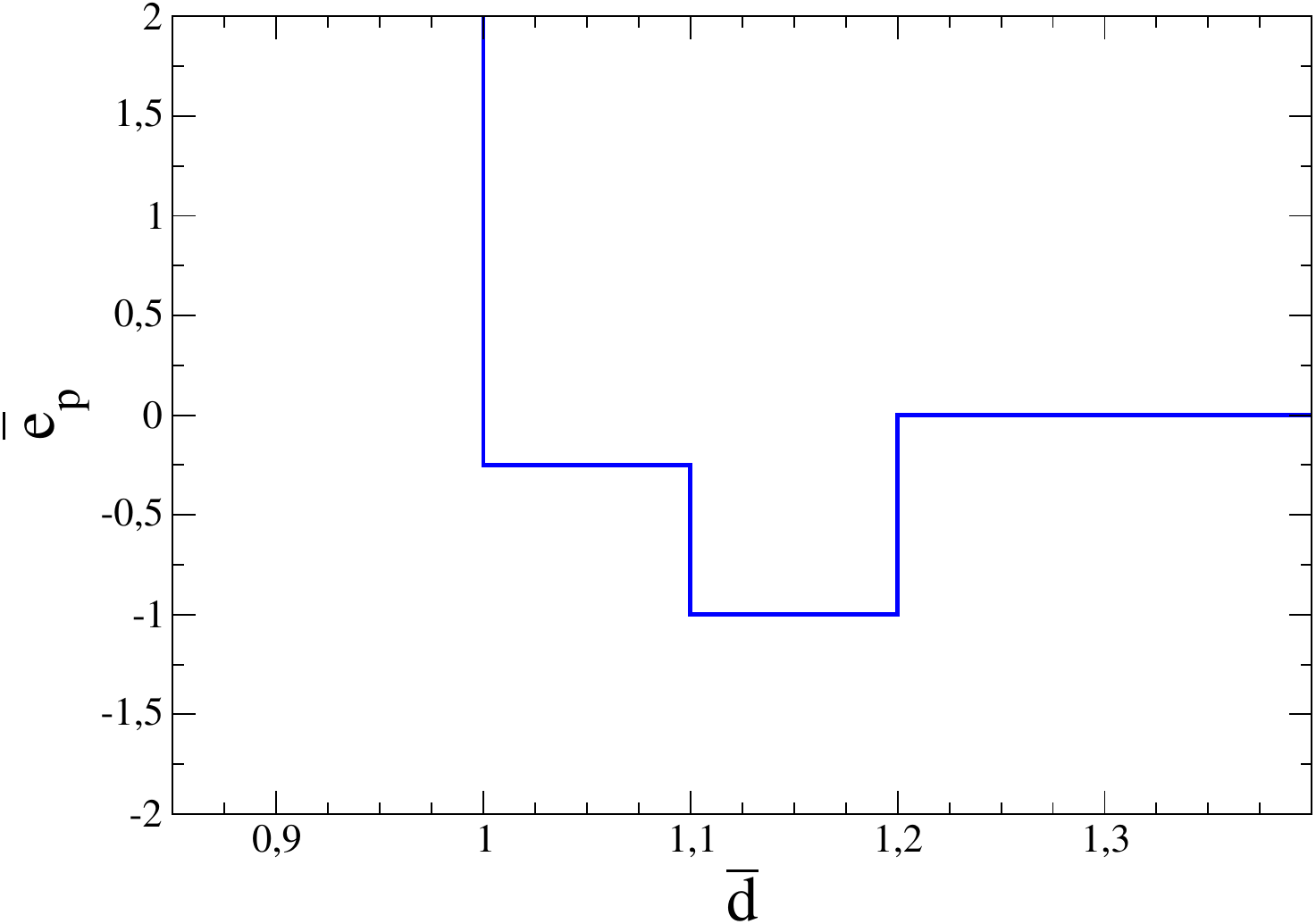}
\caption{Effective pair potential for the Associated Lattice Gas Model.}
\label{fig:7}
\end{center}
\end{figure}

This model consist as a two-scale interaction potential between water molecules. In this way, the two length scales represent 
the interaction between two neighboring particles that can form hydrogen bonds (greater distance) or not (closer distance)~\cite{Henriques2005,Szortyka2007,Szortyka2009,Fonseca2019}. In the LDL phase, the particles have an average 
distance of $\bar{d_{LDL}}=\rho^{1/2}=2/3^{1/2}$, with an average energy per particle pair of $ep_{LDL}=E_{LDL}/\rho_{LDL}=-v$. 
In the high density particles have a distance of $\bar{d_{HDL}}=\rho^{1/2}=1$ with energy per pair of particles of 
$ep_{HDL}=E_{HDL}/\rho_{HDL}=-v+2u/3$~\cite{Fonseca2019}. For the case $u=v=1$, seen in figure~\ref{fig:7}, 
the lattice leads to a hard core~\cite{Fonseca2019}.

\begin{figure}[H]
\begin{center}
\includegraphics[scale=0.3]{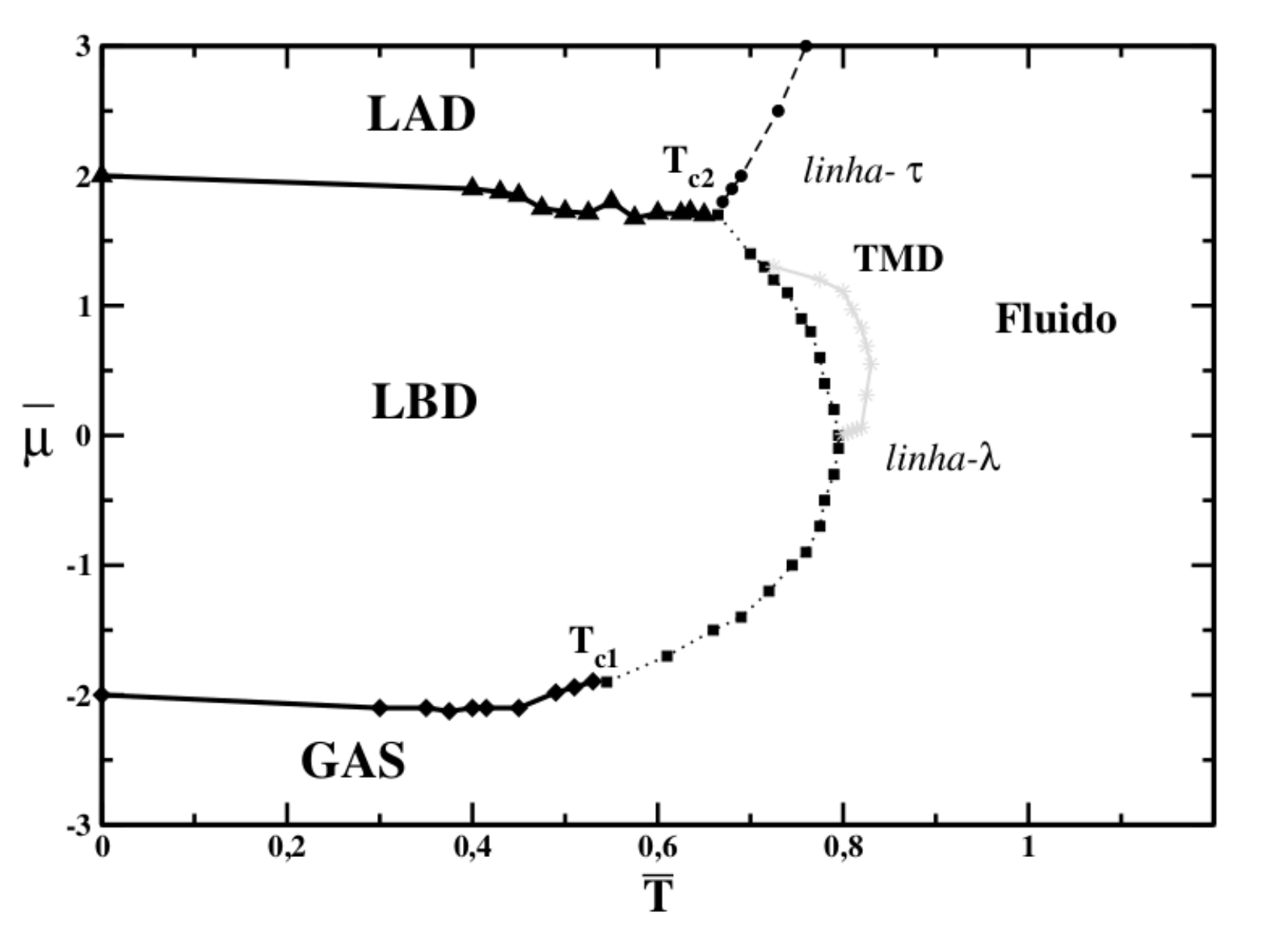} 
\caption{Reduced chemical Potential versus reduced temperature phase diagram for the bulk Associating Lattice Gas Model. 
Reproduced from reference~\cite{Szortyka2009}.}
\label{fig:8}
\end{center}
\end{figure}

A simple model that incorporates anomalous behaviors characteristic of water, such as the presence of TMD 
(the temperature of maximum density)~\cite{Henriques2005,Szortyka2007,Szortyka2009} and a region at temperatures with anomaly 
in diffusion~\cite{Henriques2005}. It has been suggested that the two length scales present in this simple model may capture 
part of the mechanism behind this set of anomalies~\cite{Fonseca2019}. Figure~\ref{fig:8} brings the properties of the ALGM 
model to the volume, where the coexistence between the phases, GAS-LDL and LDL-HDL ends in criticality~\cite{Szortyka2009}, 
with the liquid phase expected to liquid water, as well as critical lines~\cite{Fonseca2019}.

For the bulk, the system exhibits mobility~\cite{Szortyka2009} and non-zero diffusion for a wide range of low temperatures, and therefore, 
the LDL and HDL are considered liquid phases~\cite{Fonseca2019}.

We evaluated whether the entropic effects that arise, in principle, when a system is confined significantly influence when the fluid is under 
hydrophilic confinement to the point of causing a shift in the critical liquid-gas and liquid-liquid temperatures and the emergence of new structures. 
Throughout the discussion, we will highlight the differences implied by hydrophilic confinement compared to hydrophobic confinement~\cite{Fonseca2019} 
and the previously explored unconfined system~\cite{Henriques2005, Szortyka2007,Szortyka2009}.

\section{Results}

\begin{figure}[H]
\begin{center}
\includegraphics[scale=0.2]{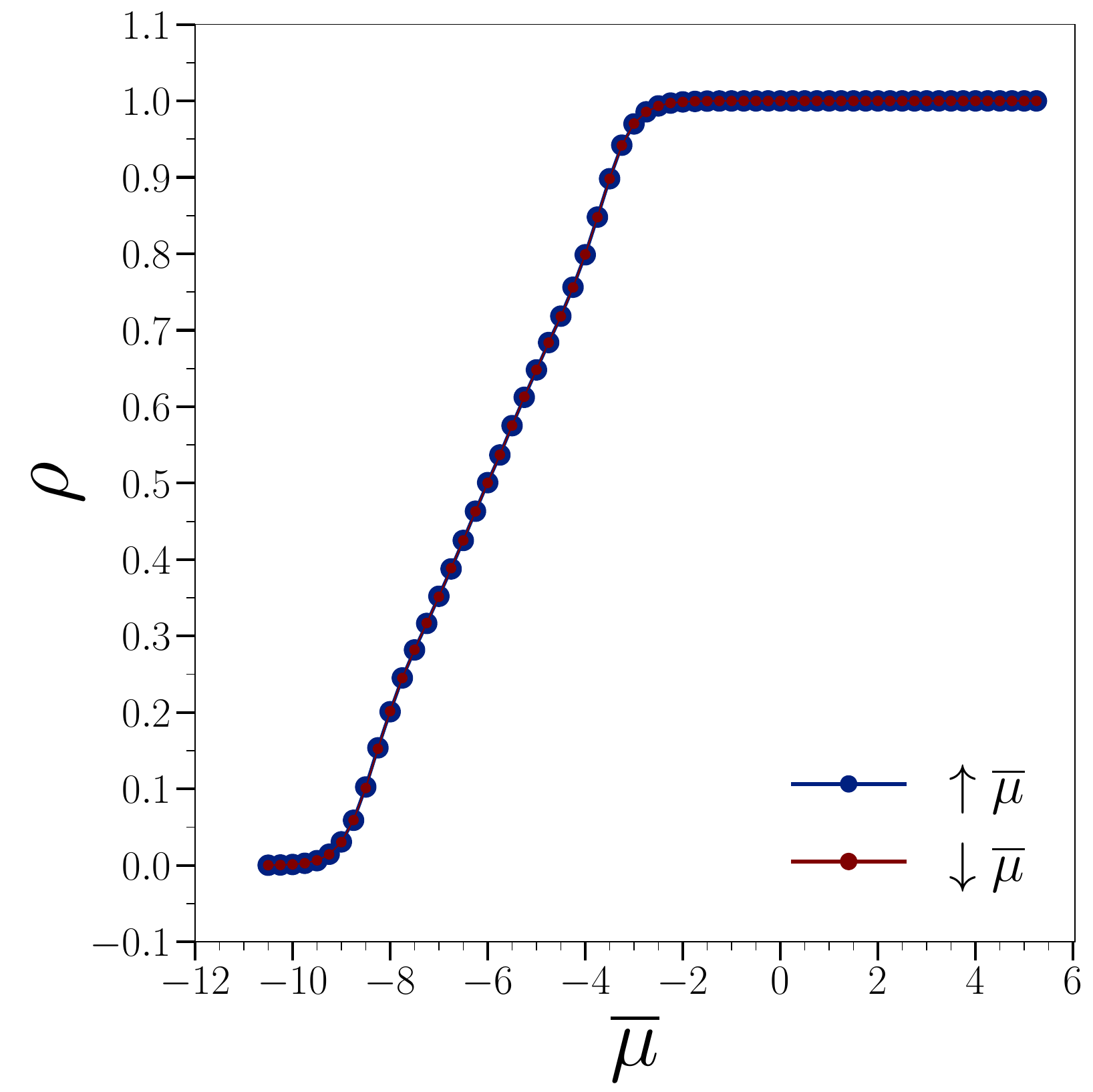}
\caption{ Density versus reduced chemical potential for $L_x$=~2 and ~$\overline{T}$=~0.30. }
\label{fig:9}
\end{center}
\end{figure}

The figure~\ref{fig:9}, shows density as a function of the reduced chemical potential of system $L_x$=~2, with two layers 
of particles type water under confinement hydrophilic. Analysis of the system under higher degree of simulated confinement, 
shows that for all investigated temperatures ($\overline{T}$=~0.30 to 0.45), when the confinement is highly hydrophilic, no phase 
transition is observed. Additionally, the LDL phase dont form, and no new structure appeared with the variation of the chemical potential. 
Not surprisingly, since that system is quasi-one-dimensional, and one-dimensional systems do not show phase transitions for short-range interactions. 

This degree of confinement still causes the gaseous phase to appear for a lower range of chemical potentials, 
occurring up to $\overline{\mu}\!\cong$~-9 and the HDL phase to appear at lower chemical potentials, thus leading to a 
greater occurrence of this phase along the variation of the chemical potential. Here the attractive wall functions as a 
higher local chemical potential, contributing to an increase in the global chemical potential of the system. 
Thus, the lattice is filled with water molecules at significantly lower chemical potentials when compared to the bulk system. 
This fact goes against the hydrophobic case in that for the same lattice size, the GAS phase persists at higher chemical potentials, 
and the HDL phase almost does not change to the duration of its occurrence about the chemical potential~\cite{Fonseca2019}.

%
%
\begin{figure}[H]
\hspace{-0.415cm}
\subfigure{%
\begin{overpic}[scale=0.23,unit=1mm]{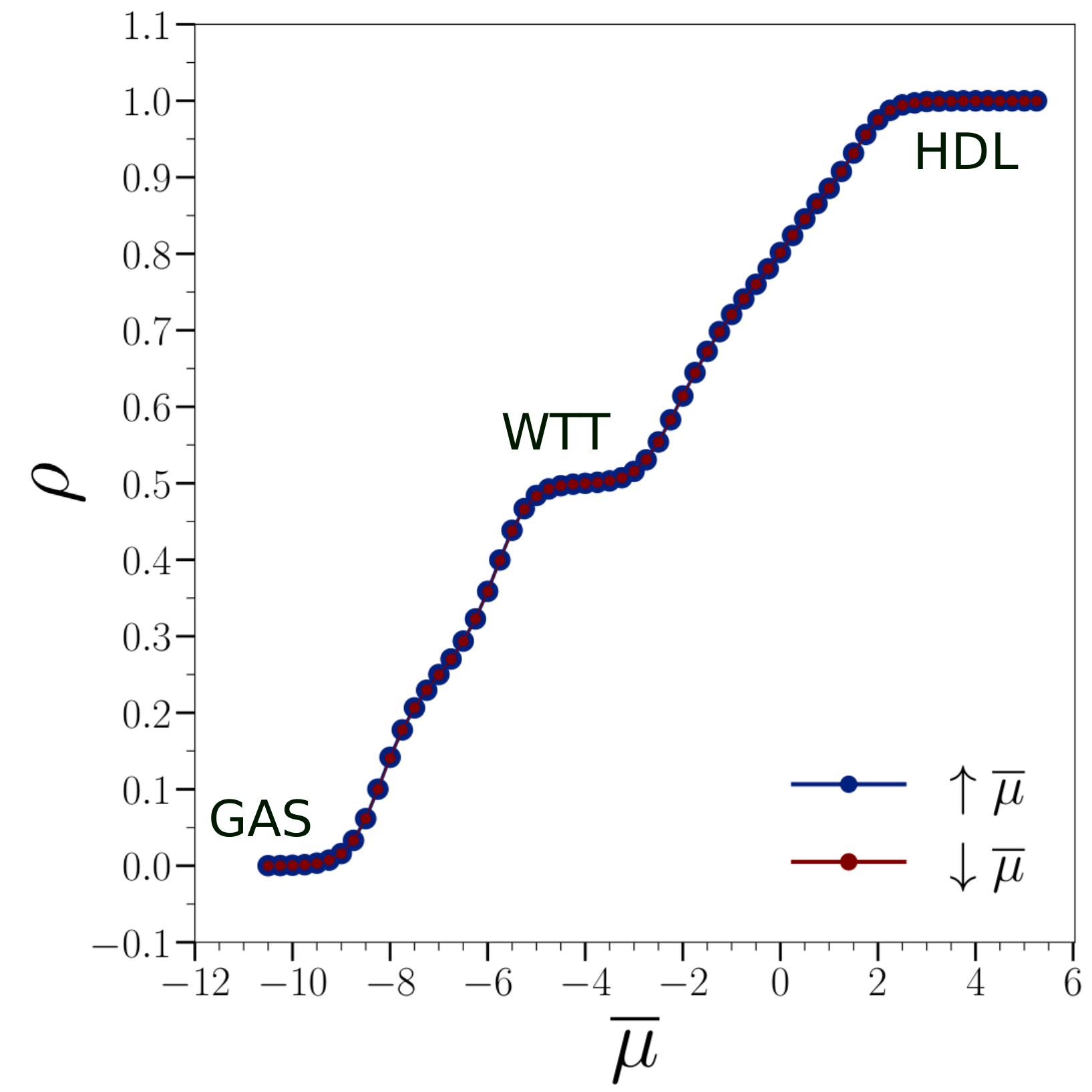}
\put(13.5,57.5){{\parbox{0.4\linewidth}{
(a)
}}}
\end{overpic}}\hfill
\subfigure{%
\begin{overpic}[scale=0.23,unit=1mm]{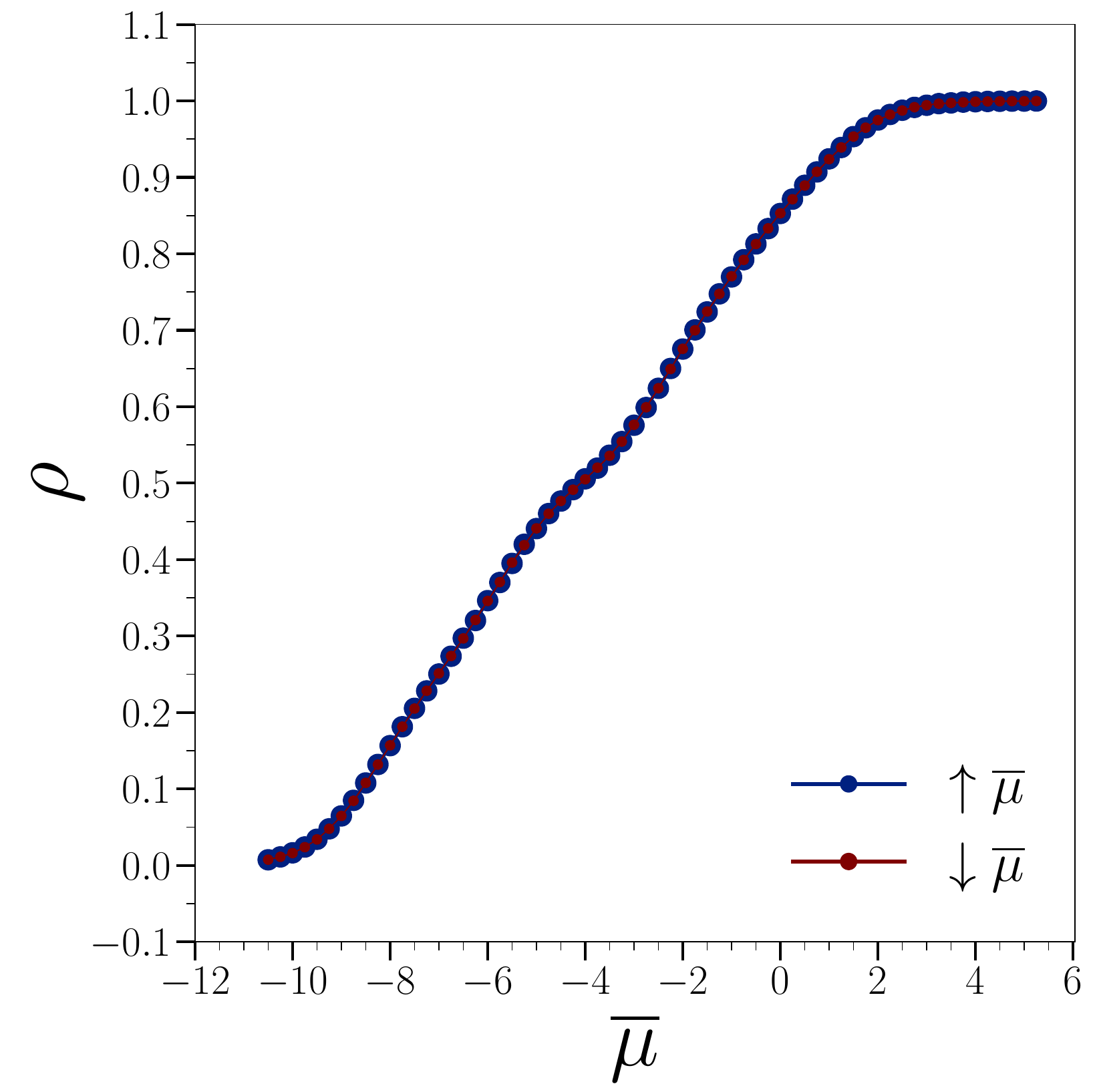}
\put(13.5,57.5){{\parbox{0.4\linewidth}{
(b)
}}}
\end{overpic}}
\caption{ Density versus reduced chemical potential for $L_x$=~4: in (a)~$\overline{T}$=~0.30 and (b)~$\overline{T}$=~0.60. }
\label{fig:10}
\end{figure}

Inspection of larger systems should reveal the degree to which hydrophilic confinement influences the behavior of the confined fluid. 
The phase diagram for $L_x$=~4, figure~\ref{fig:10}, shows The presence of the GAS phase at low chemical potentials, 
up to $\overline{\mu}\!\cong$~-9, and the HDL phase for $\overline{\mu}$ higher, from $\overline{\mu}\!\cong$~2. 
Noticed that as the chemical potential increases, the lattice full empty is gradually filled starting from the contact 
layers close to the hydrophilic matrix. The interaction between fluid particles and the wall is more attractive than the interaction between fluid particles. 
Therefore, it is energetically more favorable to the system that the water molecules form hydrogen bonds with the wall particles. 
At the beginning of the filling of the lattice, when each water molecule of the contact layers makes two hydrogen bonds 
with the wall particles, due to an angulation issue, not possible to form hydrogen bonds between fluid neighbors present in these layers. 
Thus, at low chemical potentials ($\overline{\mu}$ between -7.5 and -6.5), the sites of the contact layers are preferably 
occupied alternately to avoid the energetic penalty of having water neighbors without forming hydrogen bonds among themselves. 
A photograph of this region is shown in figure 11 -(a).
%
%
A new structure appears at still lower chemical potentials, between $\overline{\mu}$~$\!\cong$~-5.25 and $\overline{\mu}$~$\!\cong$~-3, 
with $\rho$=~0.5. The photograph of the system at $\overline{\mu}$=~-4, figure 11 -(b), shows this new structure, where the layers 
in contact with the walls are occupancy, and the GAS phase establish in the central layers. Due to this configuration, with layers immediately 
at contact with the walls filled with water molecules, there is the wetting of the contact layers, and we call this phase of ``\textit{wetting}'' (WTT). 

\begin{figure}[H]
\begin{center}
\subfigure{%
\begin{overpic}[scale=0.95,unit=1mm]{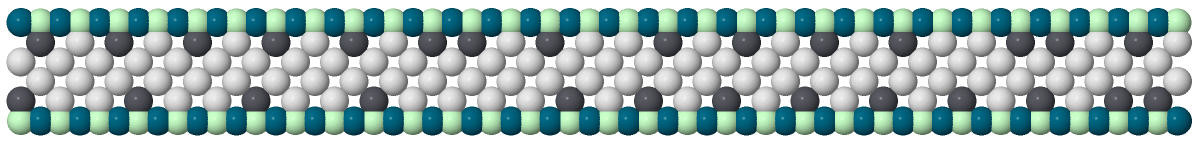}
\put(-4.75,7.0){{\parbox{0.4\linewidth}{
(a)
}}}
\end{overpic}}\hfill
\subfigure{%
\vspace{1.25cm}
\begin{overpic}[scale=0.95,unit=1mm]{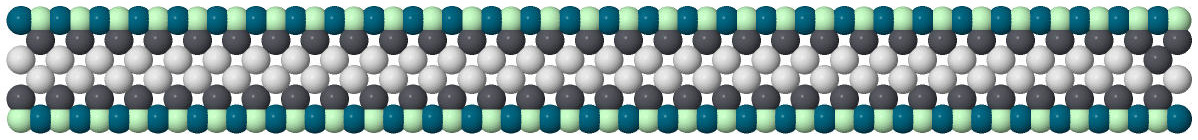}
\put(-4.75,7.0){{\parbox{0.4\linewidth}{
(b)
}}}
\end{overpic}}\hfill
\subfigure{%
\vspace{1.25cm}
\begin{overpic}[scale=0.95,unit=1mm]{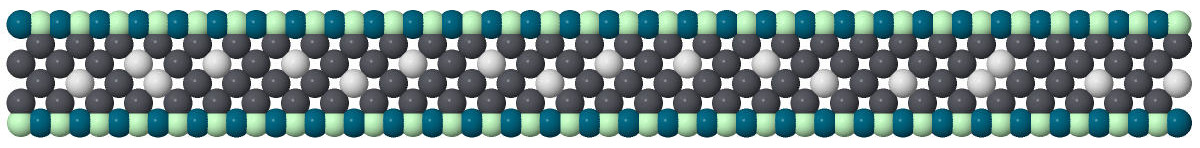}
\put(-4.75,7.0){{\parbox{0.4\linewidth}{
(c)
}}}
\end{overpic}}
\caption{ Snapshot of the 4x30 confined lattice in the x-y plane at $\overline{T}$=~0.30: in (a)~$\overline{\mu}$=~-7.0, 
(b)~$\overline{\mu}$=~-4.0 and (c)~$\overline{\mu}$=~0.75. }
\end{center}
\label{fig:11}
\end{figure}
The emergence of this ``wetting'' phase is explained by the fact that the attractive walls, in the hydrophilic confinement of our model, 
function as a higher local chemical potential, leading to the occupation of sites in the contact layers even at chemical potentials low, 
where the GAS phase would be expected to occur, compared to the unconfined system. The formation of hydrogen bonds between the lattice particles 
is energetically favorable to the system.
 
No phase transition is observed between the gas and \textit{wetting} phases, and between \textit{wetting} and high-density liquid, 
but rather a continuous passage between these structures, indicating that confinement suppresses this transition or that it occurs for 
lower temperatures, and which would require long simulations. If there is any transition, it occurs at a temperature lower than the temperatures studied for this system.

With the increase in temperature and consequent decrease in the average number of hydrogen bonds per site, we noticed the disappearance of 
the new structure, WTT, and the continuous passage from the GAS phase to the HDL phase, as can be seen in the figure~\ref{fig:10}.
%
%

%
\begin{figure}[H]
\subfigure{%
\begin{overpic}[scale=0.23,unit=1mm]{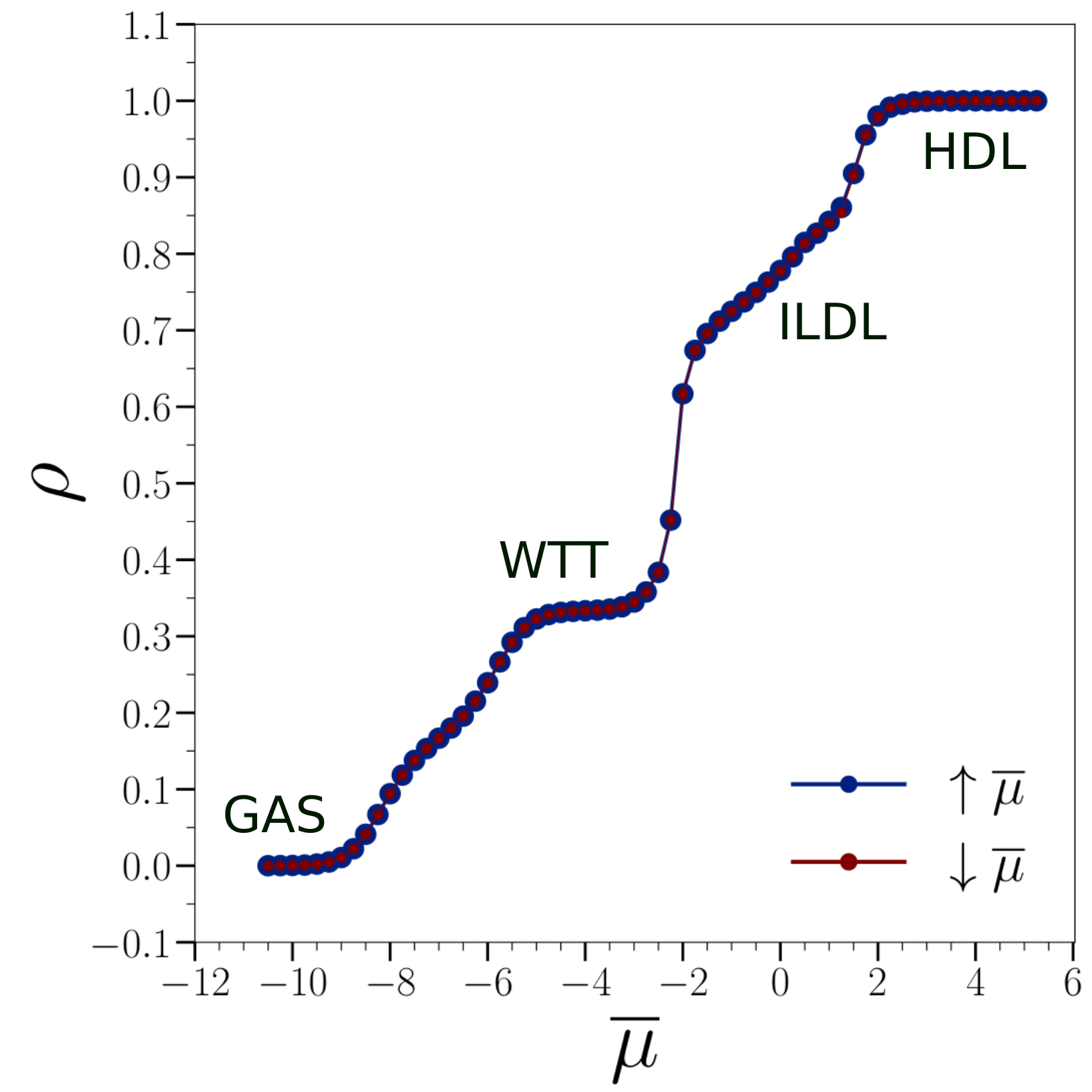}
\put(13.5,57.5){{\parbox{0.4\linewidth}{
(a)
}}}
\end{overpic}}\hfill
\subfigure{%
\begin{overpic}[scale=0.23,unit=1mm]{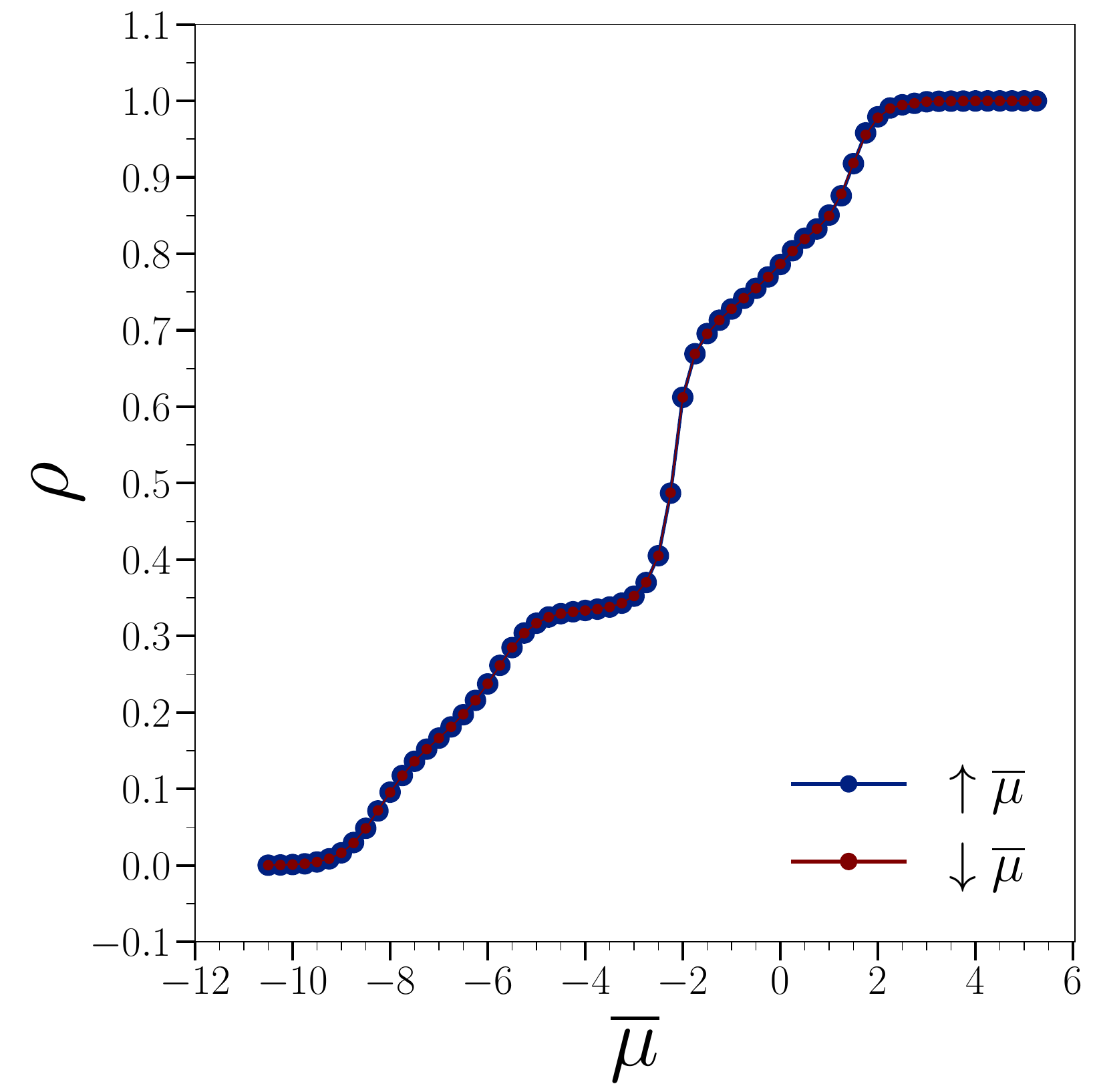}
\put(13.5,57.5){{\parbox{0.4\linewidth}{
(b)
}}}
\end{overpic}}\hfill
\subfigure{%
\begin{overpic}[scale=0.23,unit=1mm]{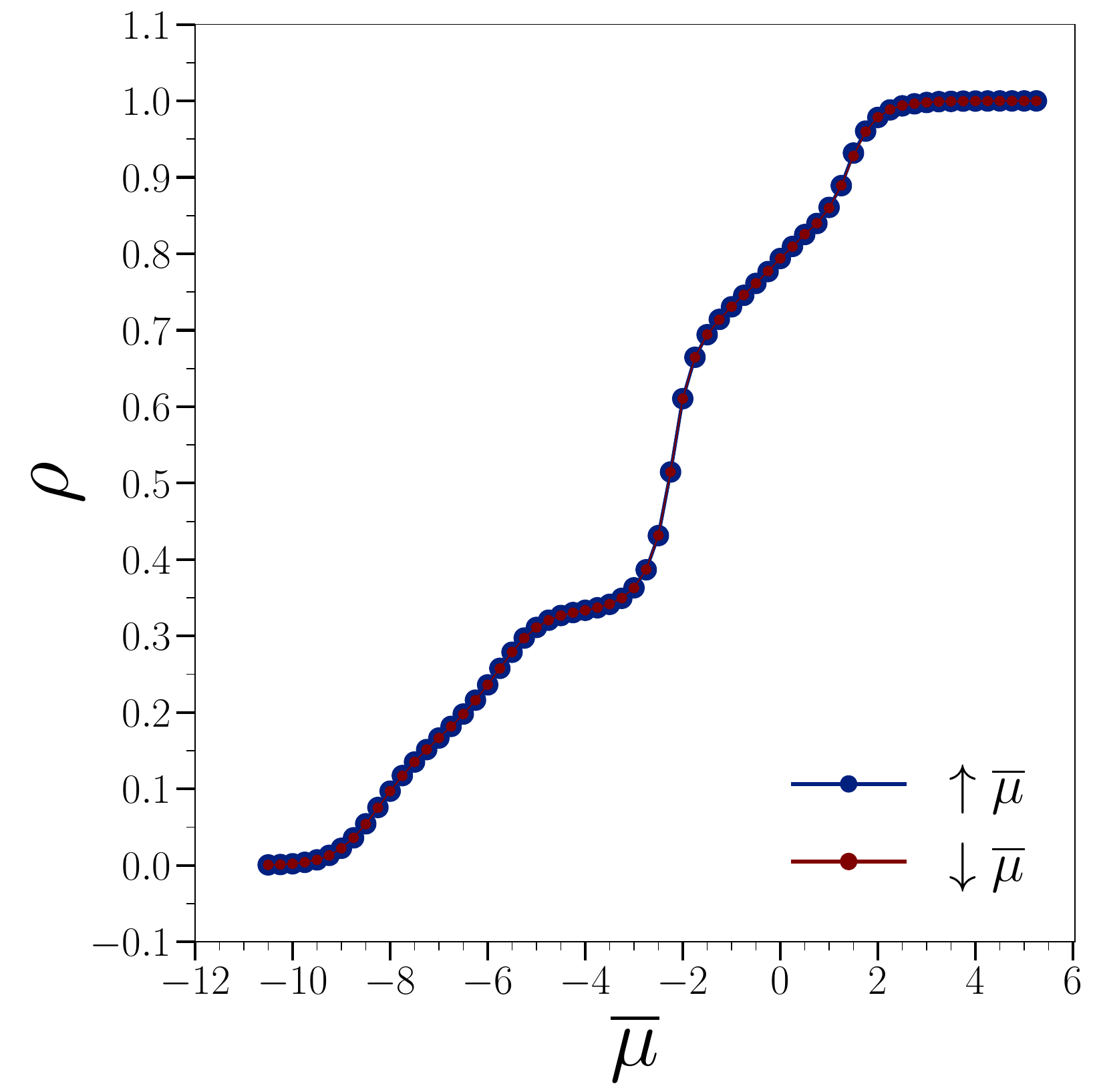}
\put(13.5,57.5){{\parbox{0.4\linewidth}{
(c)
}}}
\end{overpic}}\hfill
\subfigure{%
\begin{overpic}[scale=0.23,unit=1mm]{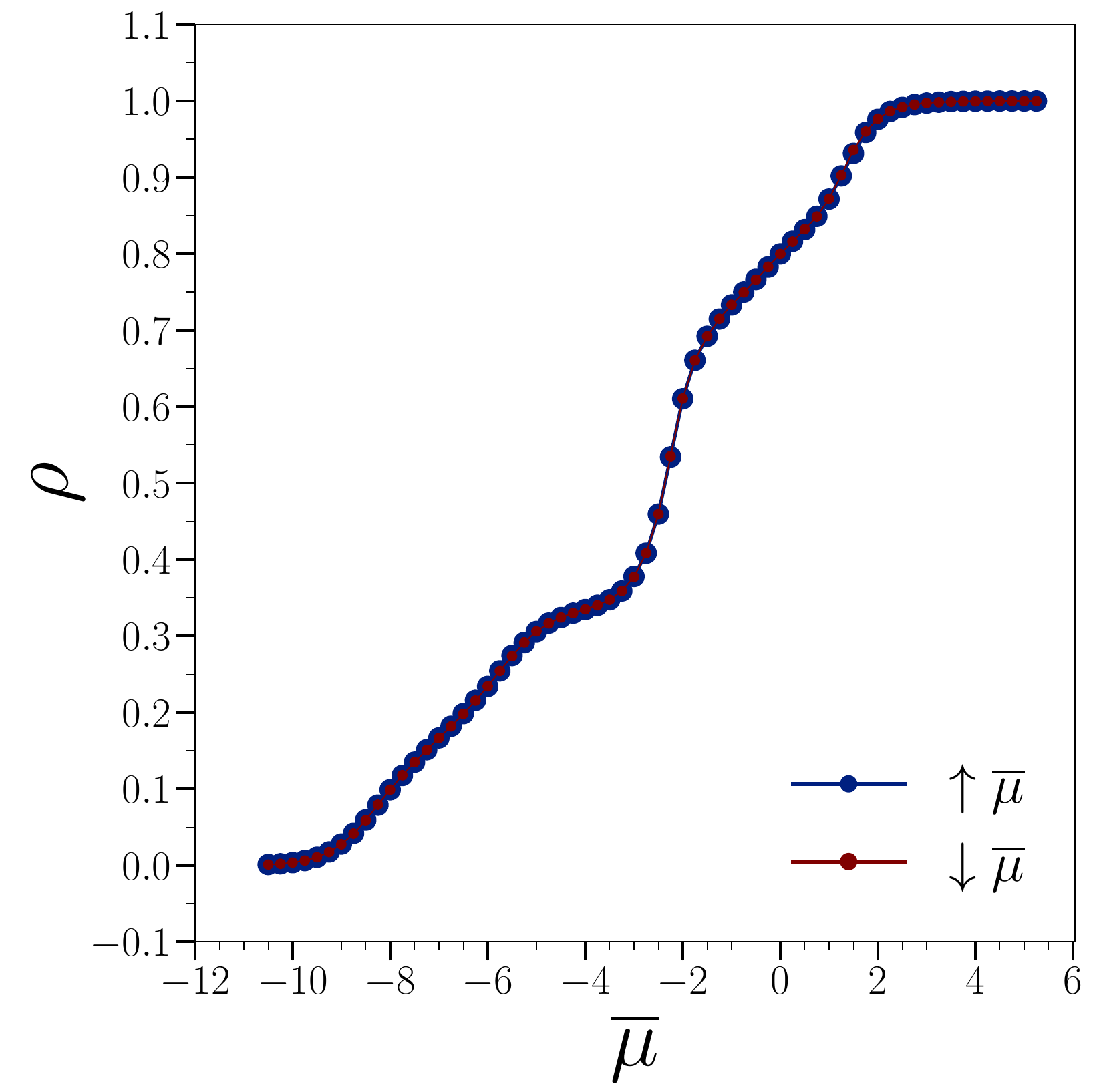}
\put(13.5,57.5){{\parbox{0.4\linewidth}{
(d)
}}}
\end{overpic}}
\caption{ Density versus reduced chemical potential for $L_x$=~6 and (a)~$\overline{T}$=~0.30, (b)~$\overline{T}$=~0.35, (c)~$\overline{T}$=~0.40 and (d)~$\overline{T}$=~0.45. }
\label{fig:12}
\end{figure}

\begin{figure}[H]
\begin{center}
\includegraphics[scale=0.95]{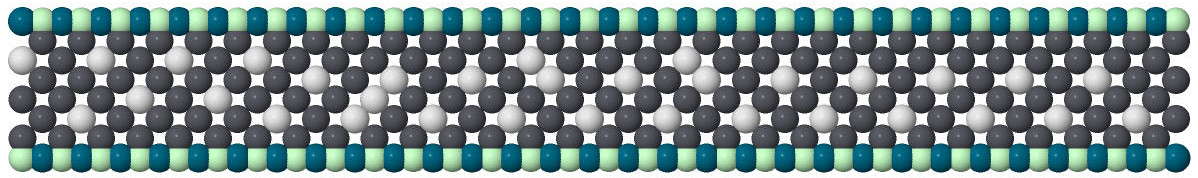}
\caption{ Photo of the 6x30 confined lattice in the x-y plane at $\overline{T}$=~0.30 and $\overline{\mu}$=~0.75, where the ILDL phase 
(LDL with interference from the attractive walls that wet the contact layers) is shown. }
\label{fig:13}
\end{center}
\end{figure}

To observe if this new structure, \textit{wetting}, is present in other degrees of confinement, we also studied the systems $L_x$=~6, 8, 10, 12, and 14. 
Through the curves of density versus reduced chemical potential, figure \ref{fig:12}, it is easy to notice the presence 
of the WTT phase for the 6x30 lattice, with density around $\rho$=~0.33 and reduced chemical potential between $\overline{\mu} $=~-5.75 and -3.0. 
In addition to this phase, the GAS phase is present, occurring for low chemical potentials, $\overline{\mu} \leqslant$~-9, and 
the HDL phase for $\overline{\mu} \geqslant$~-2. Also notable is the structure that appears for values $\overline{\mu}$ between -1.75 and 1.25, 
with density corresponding to the range $\rho \cong$~0.67 to 0.86. In this configuration, as shown in figure \ref{fig:13}, the fluid molecules fill 
the contact layers, and the central layers particles are arranged in a structure that resembles the LDL phase. Thus, we identify this new structure as 
LDL phase with interference from the attractive walls that wet the contact layers, ILDL ``\textit{interfered LDL}'.

Many water systems under nanoscale confinements exhibit different multiphase \cite{evans2019unified,shaat2019fluidity} structures. 
The formed structures depend on the water-surface interactions, geometry, and confinement size. In particular, wetting of hydrophilic surfaces 
is often observed \cite{verdaguer2006molecular,malani2009influence}.

There is the very smooth change from the gas phase to the ``\textit{wetting}'' phase. 
Between the WTT and ILDL phases, we observe an abrupt variation in the system density, indicating a first-order transition. 
Not observed transition between the ILDL and HDL structures, but a continuous passage between these indicates that confinement 
suppresses this transition or that it occurs at lower temperatures.

When we increase the distance between the adjoining walls to $Lx$=~8 (figure 14), 
at $\overline{T}$=~0.30, the lattice is occupied continuously, since $\overline{\mu}$=~-8.75, preferably with the sites 
in contact with the attractive walls occupied alternately, until both contact layers are occupiedcompletely, around -~4.75~$\leq \overline{\mu}\leq$~-~3.25. 
At this point, the system is in the ``\textit{wetting}'' phase and transitions to the ILDL phase, and from there, for higher chemical potentials, 
it undergoes another transition, now to the HDL phase. Phases of the 8x30 lattice are in figure \ref{fig:15}. Therefore, we observe two first-order transitions 
for low values of $\overline{T}$: the transition between \textit{wetting} and low-density liquid under wall interference, WTT-ILDL, 
and the transition of this last phase for the high density liquid, ILDL-HDL.
\begin{figure*}[H]
\subfigure{%
\begin{overpic}[scale=0.3,unit=1mm]{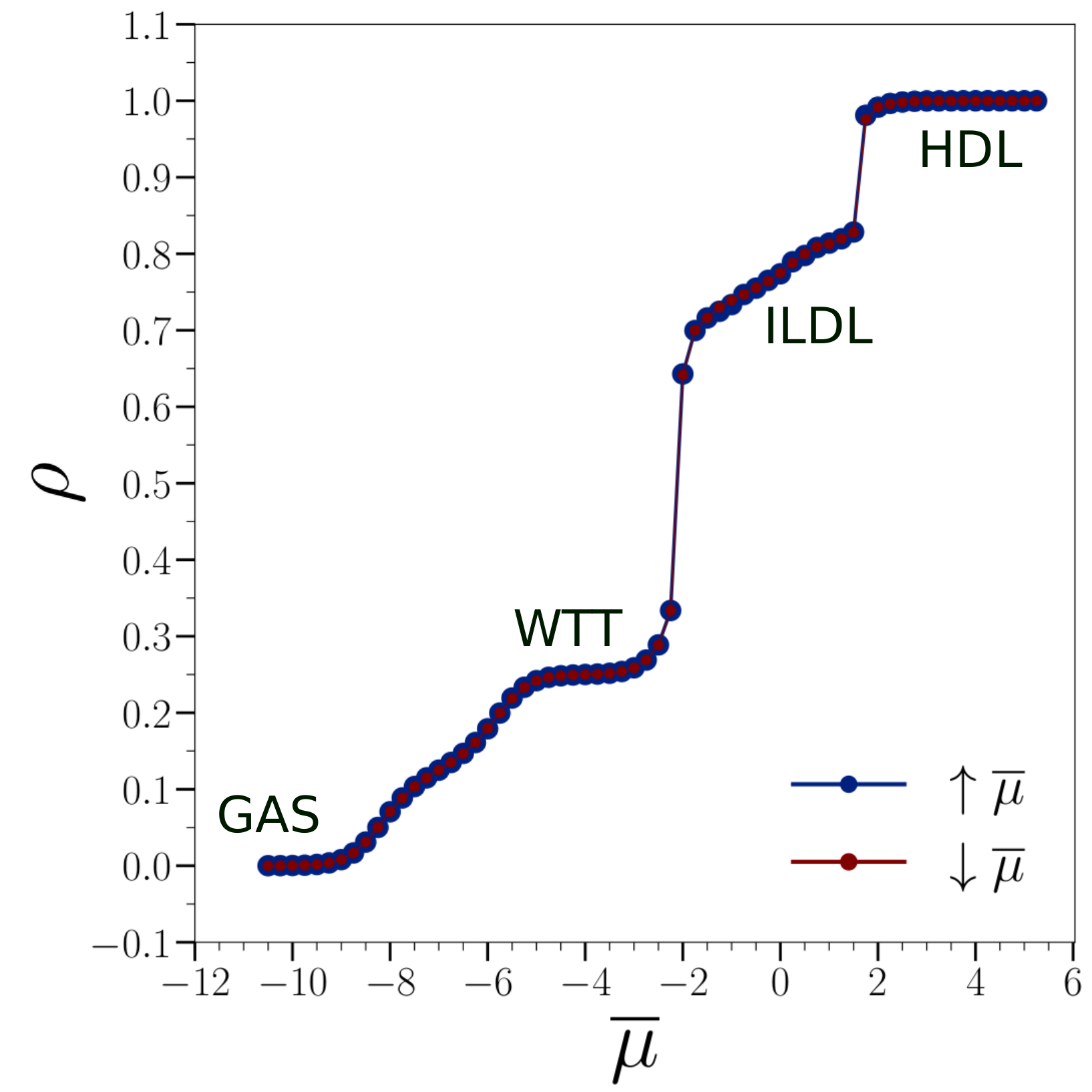}
\put(13.5,57.5){{\parbox{0.4\linewidth}{
(a)
}}}
\end{overpic}}\hfill
\subfigure{%
\begin{overpic}[scale=0.3,unit=1mm]{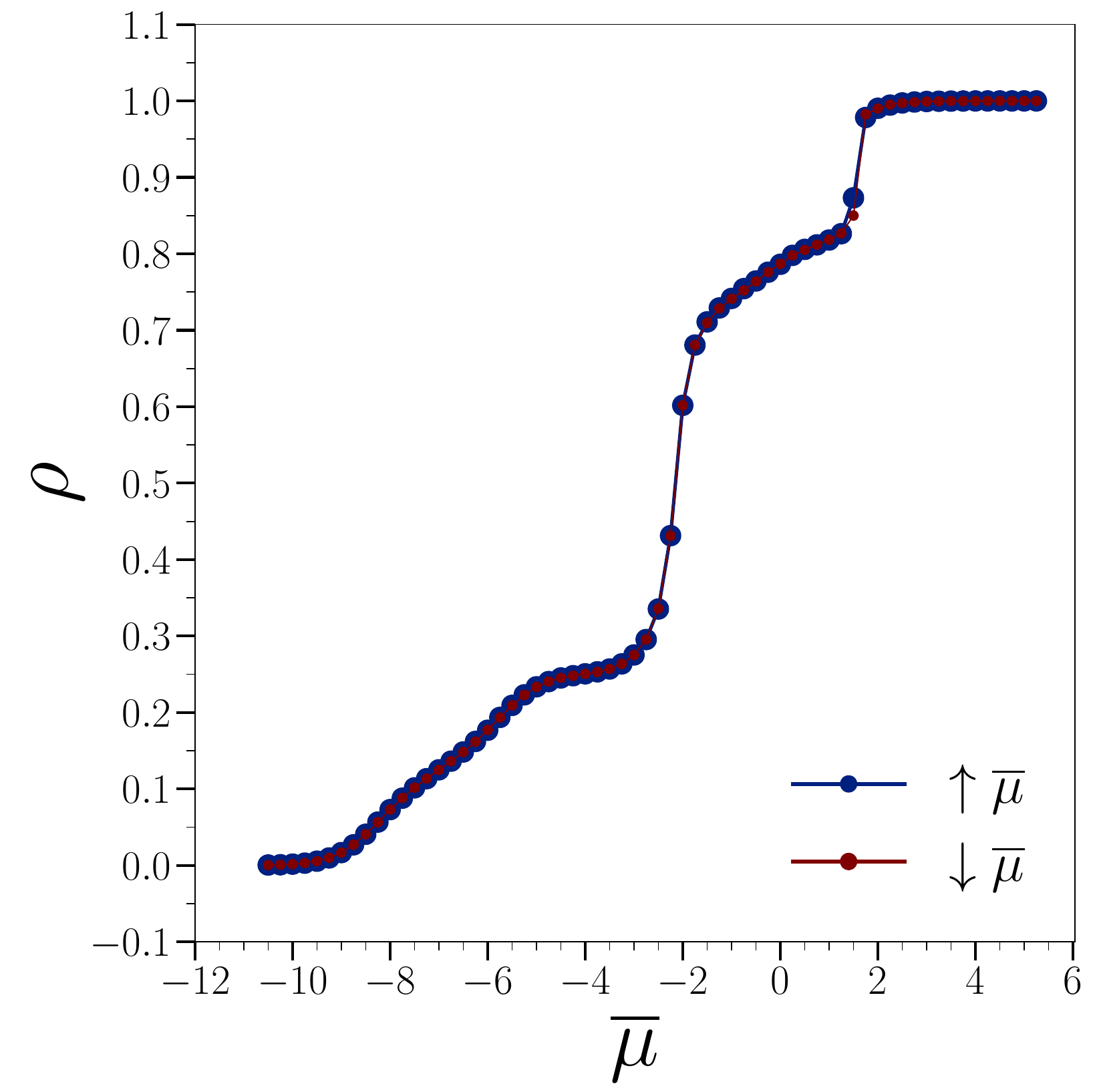}
\put(13.5,57.5){{\parbox{0.4\linewidth}{
(b)
}}}
\end{overpic}}\hfill
\subfigure{%
\begin{overpic}[scale=0.3,unit=1mm]{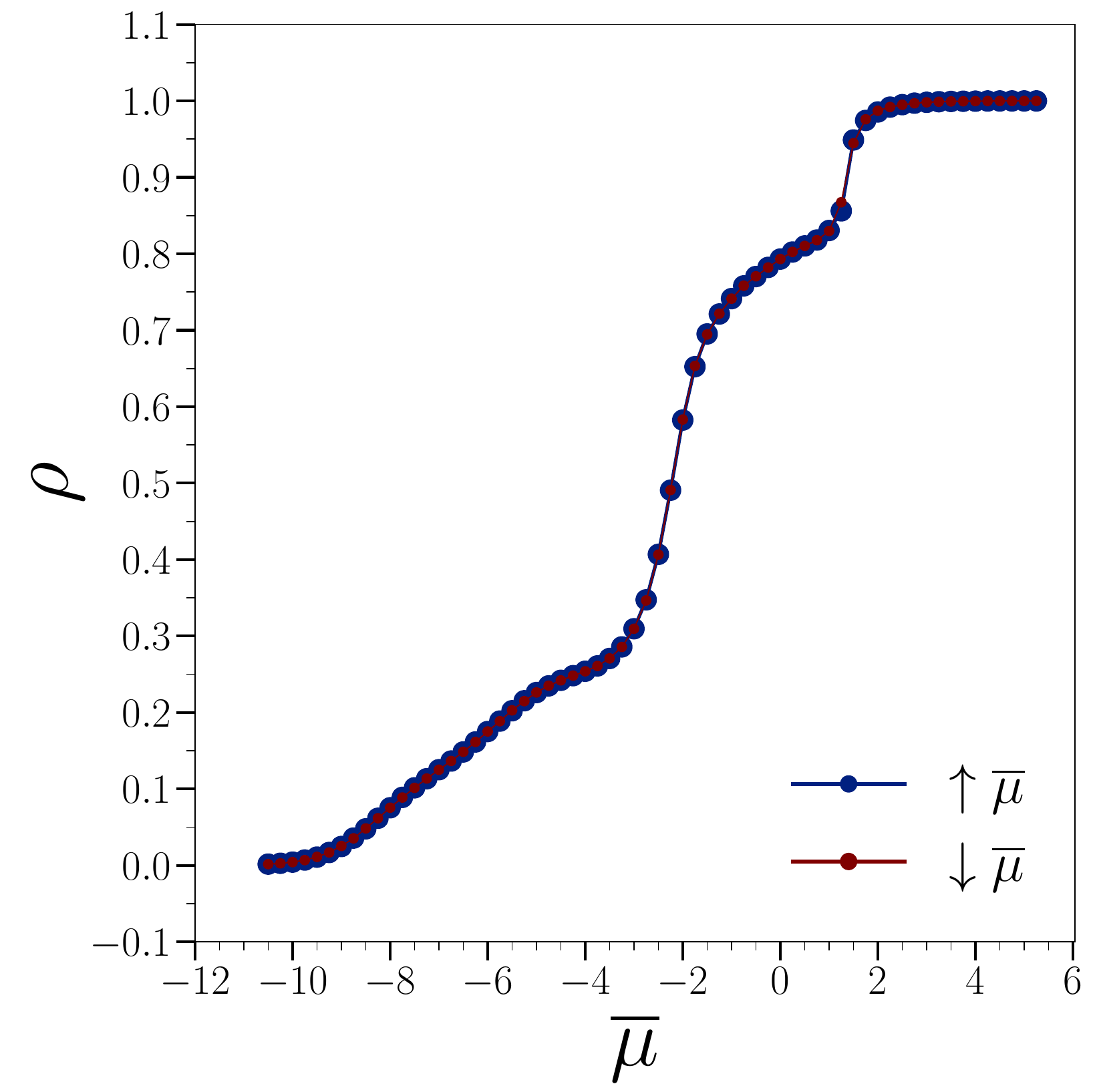}
\put(13.5,57.5){{\parbox{0.4\linewidth}{
(c)
}}}
\end{overpic}}\hfill
\subfigure{%
\begin{overpic}[scale=0.3,unit=1mm]{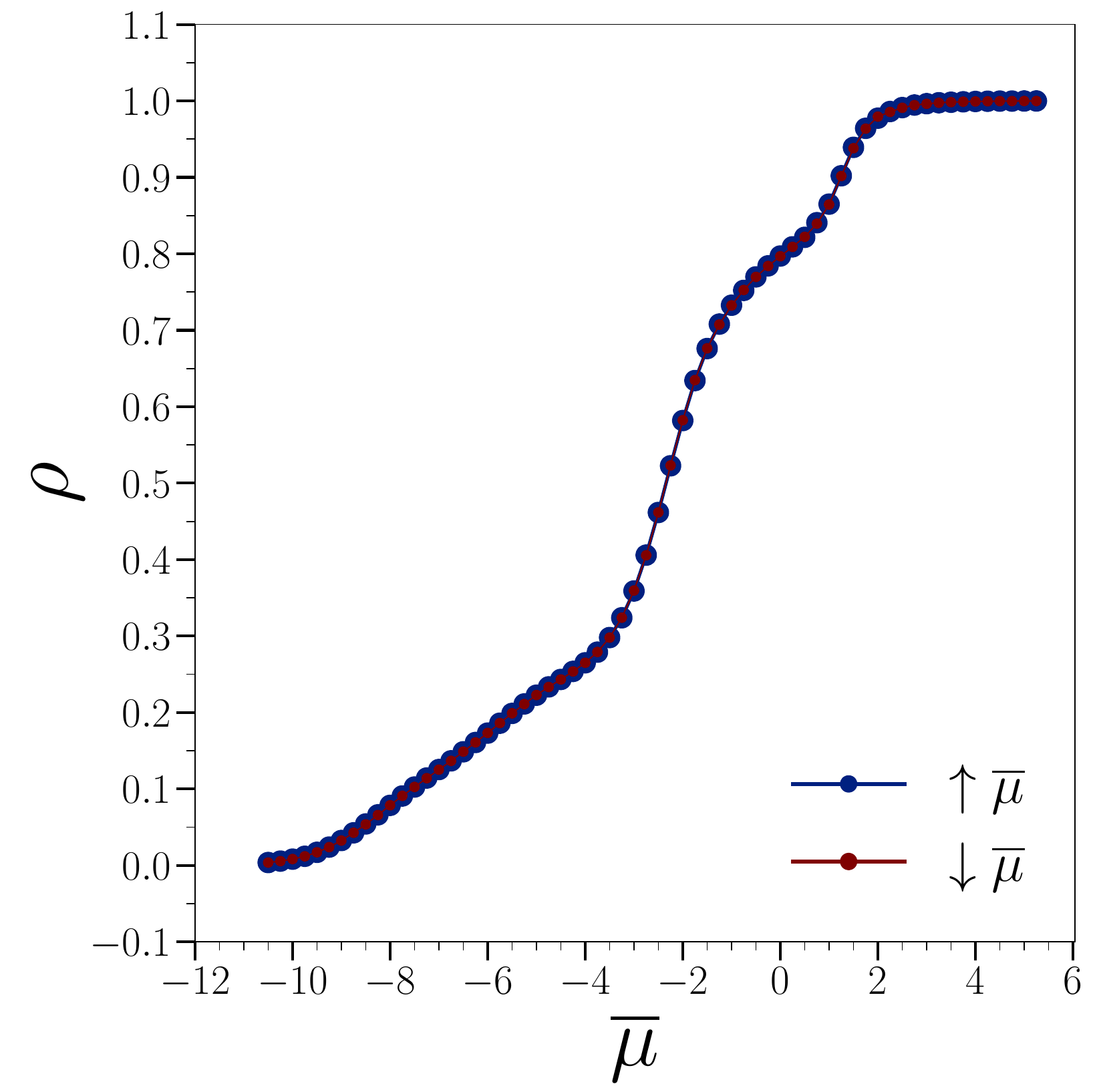}
\put(13.5,57.5){{\parbox{0.4\linewidth}{
(d)
}}}
\end{overpic}}
\caption{ Density versus reduced chemical potential for $L_x$=~8 at (a)~$\overline{T}$=~0.30, (b)~$\overline{T}$=~0.35, 
(c)~$\overline{T}$=~0.40, (d)~$\overline{T}$=~0.45, (e)~$\overline{T}$=~0.50, (f)~$\overline{T}$=~0.55 and 
(g)~$\overline{T}$=~0.60. }
\label{fig:14}
\end{figure*}
\begin{figure}[H]
\begin{center}
\subfigure{%
\vspace{1.25cm}
\begin{overpic}[scale=0.95,unit=1mm]{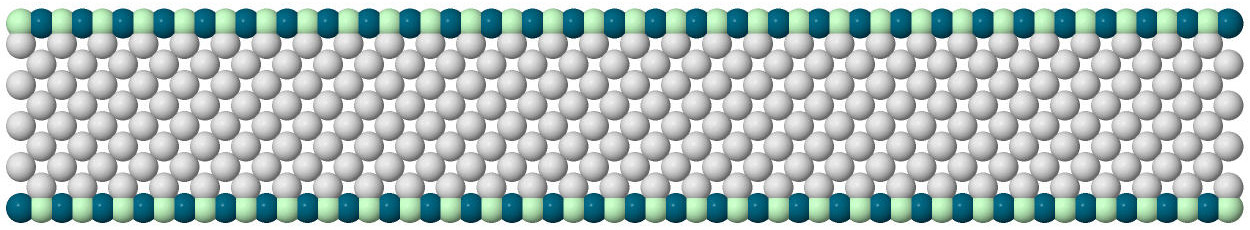}
\put(-4.75,13.0){{\parbox{0.4\linewidth}{
(a)
}}}
\end{overpic}}\hfill
\subfigure{%
\vspace{1.25cm}
\begin{overpic}[scale=1.0,unit=1mm]{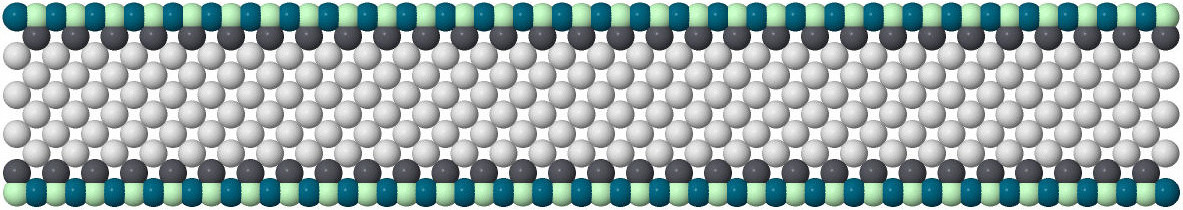}
\put(-4.75,13.0){{\parbox{0.4\linewidth}{
(b)
}}}
\end{overpic}}\hfill
\subfigure{%
\vspace{1.25cm}
\begin{overpic}[scale=1.0,unit=1mm]{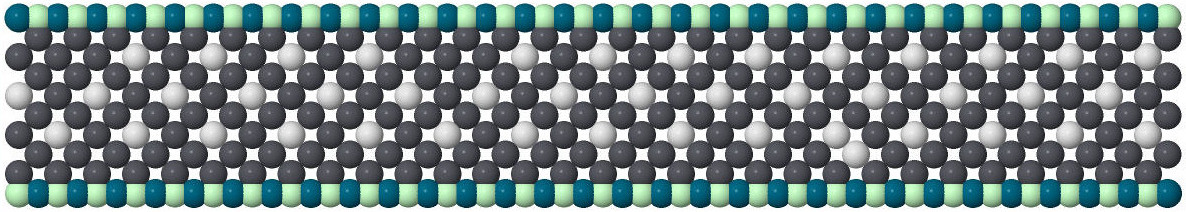}
\put(-4.75,13.0){{\parbox{0.4\linewidth}{
(c)
}}}
\end{overpic}}\hfill
\subfigure{%
\vspace{1.25cm}
\hspace{0.075cm}
\begin{overpic}[scale=0.945,unit=1mm]{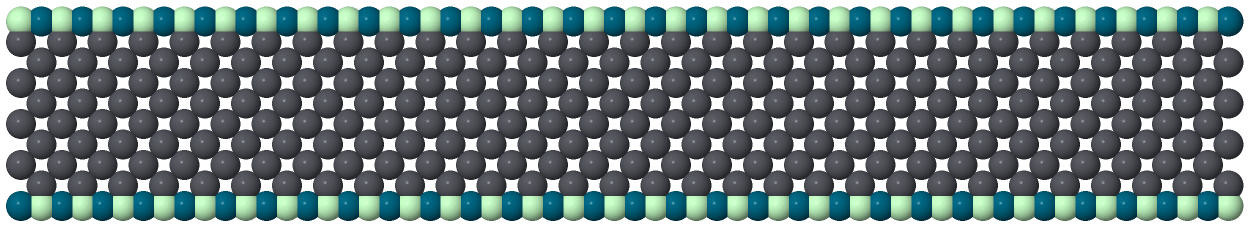}
\put(-4.75,13.0){{\parbox{0.4\linewidth}{
(d)
}}}
\end{overpic}}
\caption{ Snapshot of the 8x30 confined lattice in the x-y plane at $\overline{T}$= 0.30, showing its phases: in (a)~GAS ($\overline{\mu}$=~-10.0), 
(b)~WTT ($\overline{\mu}$=~-4.0), (c)~ILDL ($\overline{\mu}$=~0.75) and (d)~HDL ($\overline{\mu}$=~4.0). }
\label{fig:15}
\end{center}
\end{figure}

To approximately identify the critical point that ends the ILDL-HDL transition, we analyzed more temperatures 
for $L_x$=~8 and identified $\overline{T}_{(ILDL-HDL)}$=~0.50 and $\overline{\mu}_{\,(ILDL-HDL)}$=~1.63.

\begin{figure}[H]
\subfigure{%
\begin{overpic}[scale=0.23,unit=1mm]{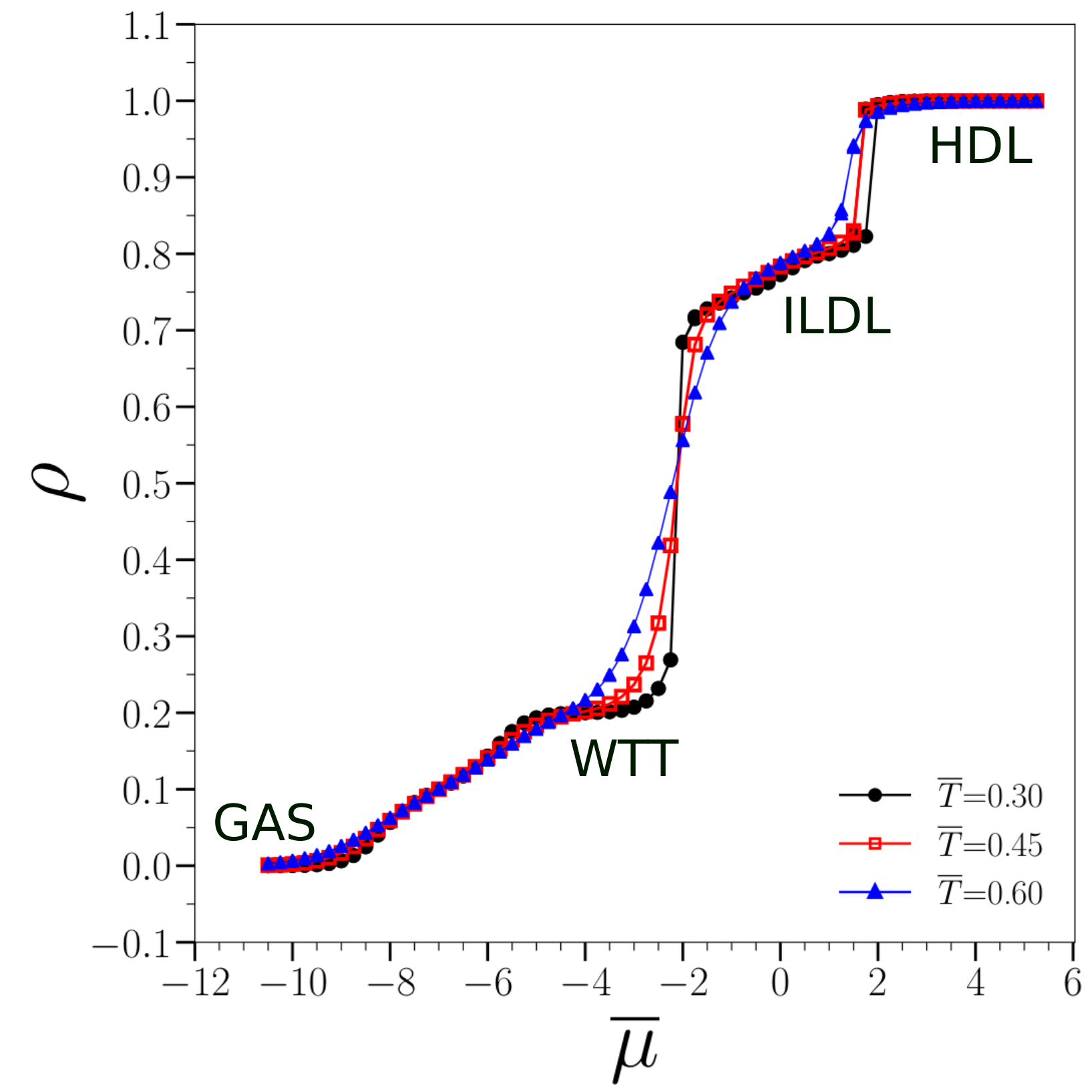}
\put(13.5,57.5){{\parbox{0.4\linewidth}{
(a)
}}}
\end{overpic}}\hfill
\subfigure{%
\begin{overpic}[scale=0.23,unit=1mm]{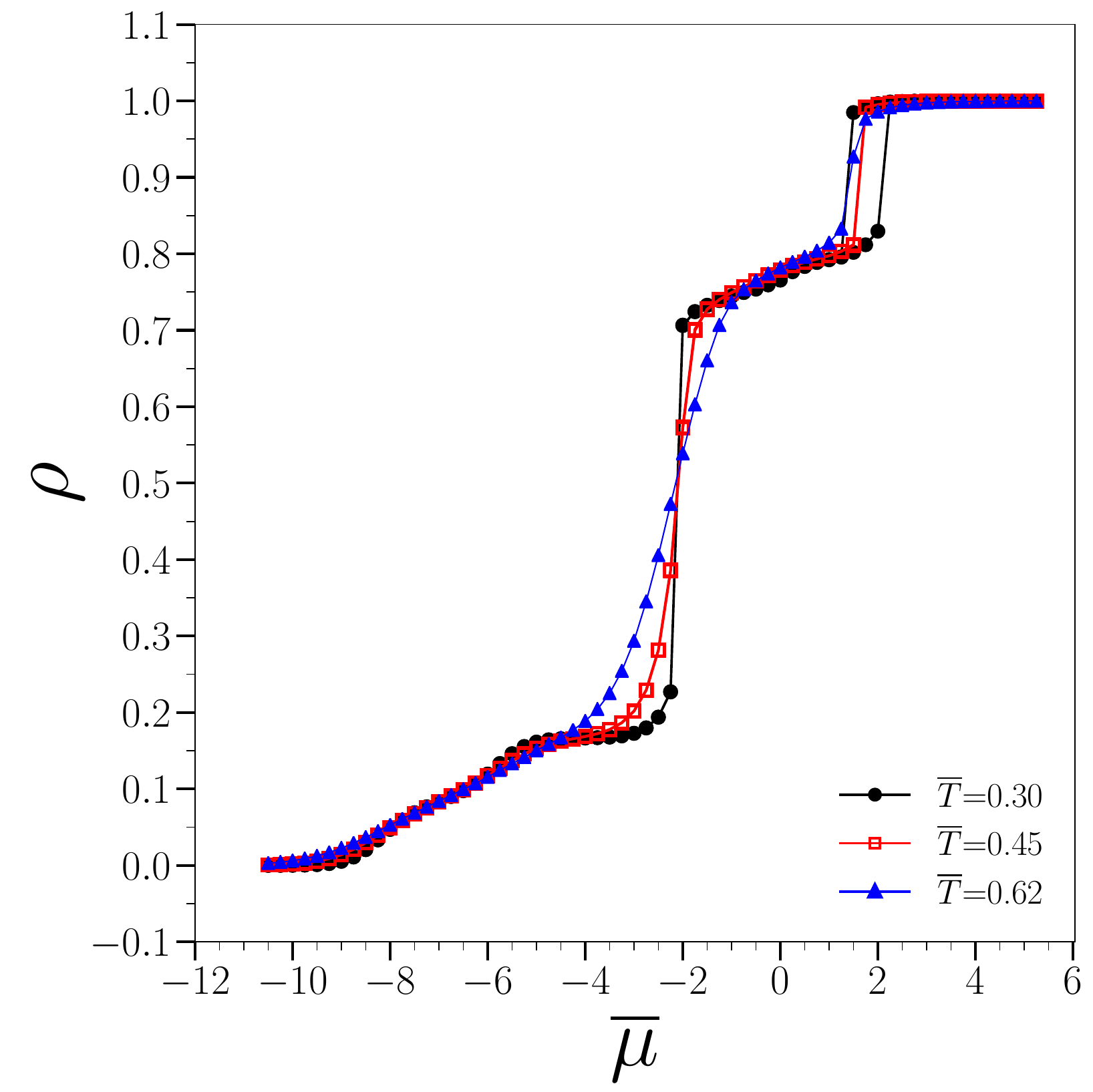}
\put(13.5,57.5){{\parbox{0.4\linewidth}{
(b)
}}}
\end{overpic}}\hfill
\subfigure{%
\hspace{3.25cm}
\begin{overpic}[scale=0.23,unit=1mm]{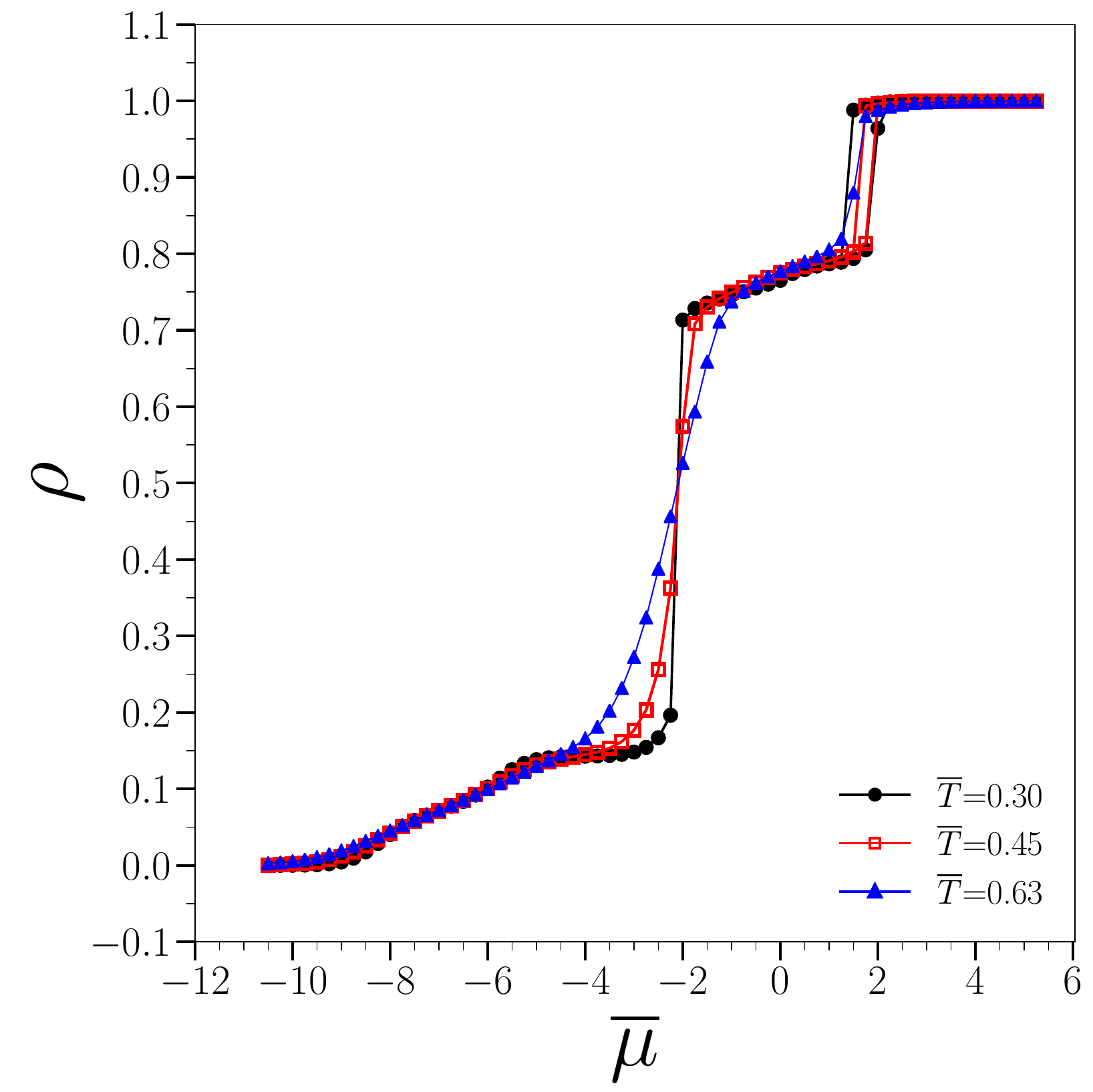}
\put(13.5,57.5){{\parbox{0.4\linewidth}{
(c)
}}}
\end{overpic}}
\caption{ Density versus reduced chemical potential for various reduced temperatures and (a)~$L_x$=~10, (b)~$L_x$=~12 e (c)~$L_x$=~14. }
\label{fig:16}
\end{figure}

Analysis of the larger systems, $L_x$=~10, 12 and 14, figure~\ref{fig:16}, 
indicates that as $L_x$ increases, the critical temperature of the ILDL-HDL transition will also increase. 
The behavior of the critical temperature as a function of the variation in the degree of confinement is explored in figure \ref{fig:19}. 
The value of $Lx$=~30, corresponding to the critical point of the bulk system, is the simulated finite value, not counting the periodic 
boundary conditions (which make the unconfined system infinite). Can see that the critical temperature of the transition ILDL-HDL increases as the 
distance between the confining walls becomes greater, but being lower than the temperature of the transition LDL-HDL in the bulk system, 
even for the largest simulated lattice size.

\begin{figure}[H]
\vspace{0.5cm}
\subfigure{%
\hspace{-0.75cm}
\begin{overpic}[scale=0.23,unit=1mm]{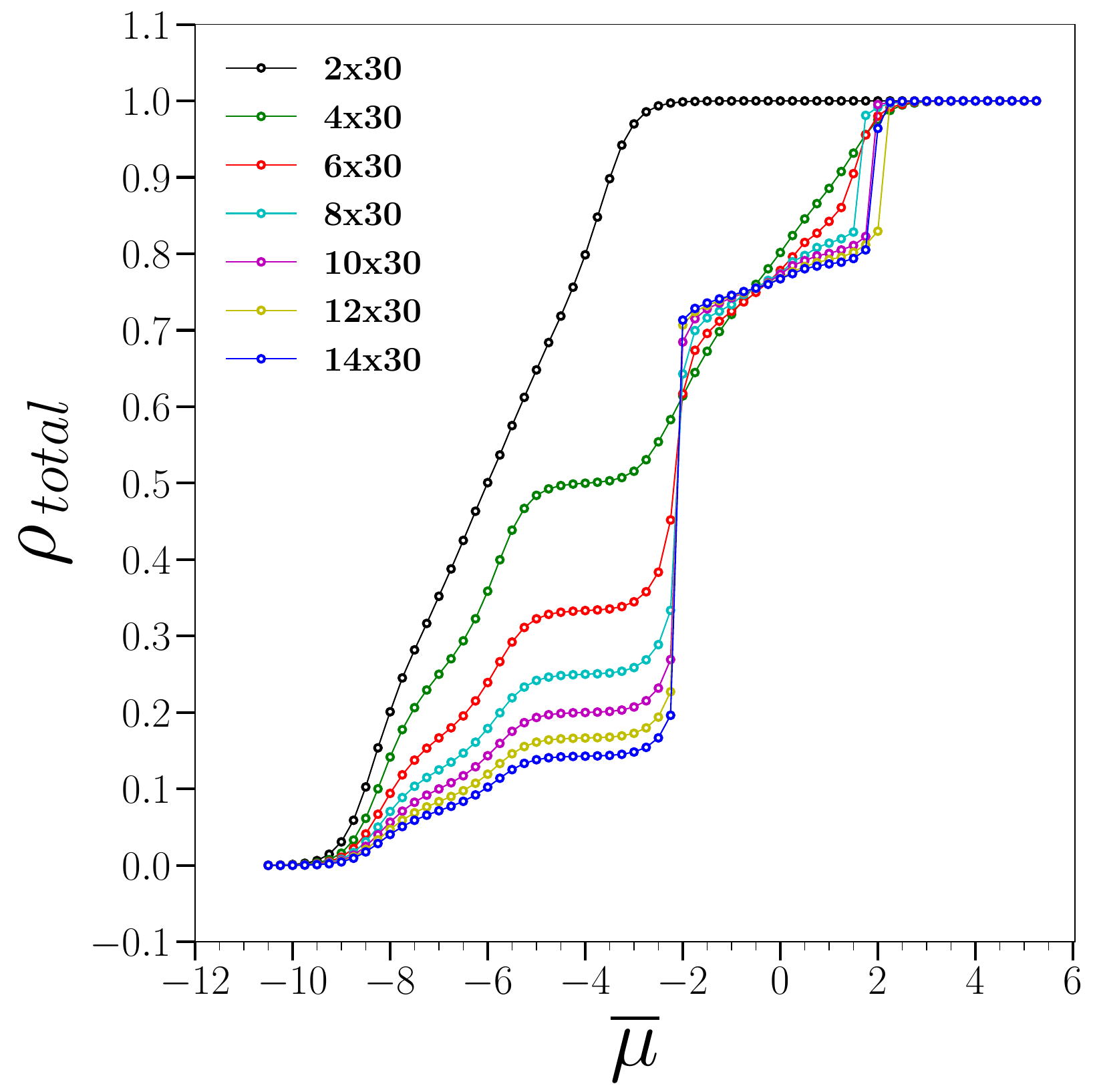}
\put(57.5,10.75){{\parbox{0.4\linewidth}{
(a)
}}}
\end{overpic}}\hfill
\subfigure{%
\begin{overpic}[scale=0.23,unit=1mm]{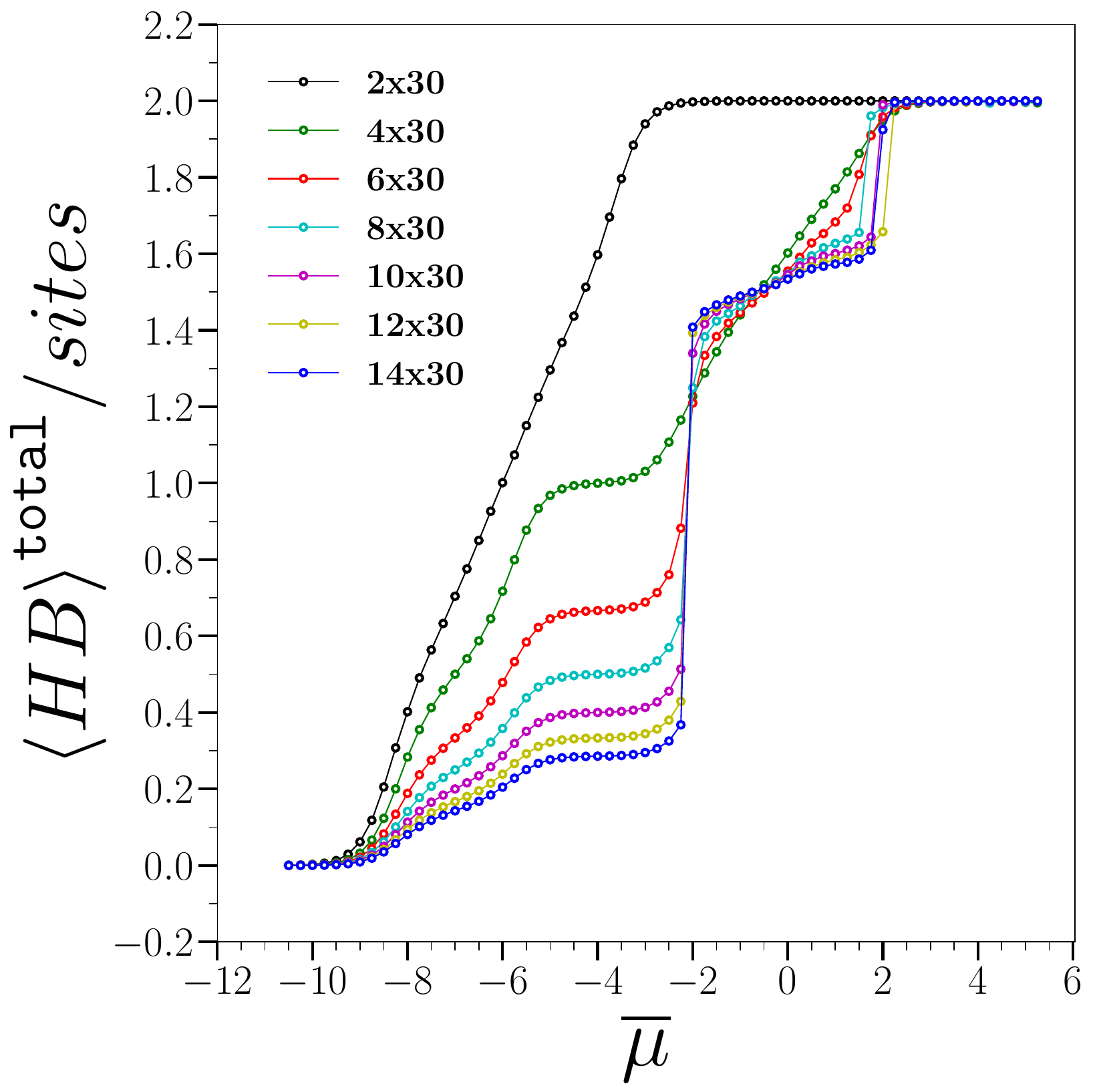}
\put(57.5,10.75){{\parbox{0.4\linewidth}{
(b)
}}}
\end{overpic}}
\caption{ Total density versus chemical potential reduced to $\overline{T}$=~0.30 for different values of $L_x$ in (a). 
And in (b): average of the total number of hydrogen bonds, fluid-fluid and fluid-wall, as a function of the total number of 
sites in the confined lattice versus reduced chemical potential for the different degrees of confinement of the system simulated to $\overline {T}$=~0.30. }
\label{fig:17}
\end{figure}

When investigating the total density of the system as a function of the chemical potential at $T$=~0.30, 
figure \ref{fig:17}~(a), we observe a different behavior between $L_x$=~2 and the other lattice sizes. 
For $L_x$=~2, the system, initially empty (GAS phase), fills up as the $\overline{\mu}$ increases, and reaches the HDL phase, 
with the lattice full for $\overline{\mu}\geq$~-3. In this case, we count the total density as the density of the contact layers. 
When the contact layers are fully occupied, the lattice is full ($\rho$=~1). For the other lattice sizes, after the complete occupation 
of the sites in the contact layers (the point at which the \textit{wetting} phase is reached), the sites in the central layers are 
occupied in an unsuccessful attempt to reach the LDL configuration of the system bulk. Due to the presence of the attractive walls, 
the ILDL phase (LDL phase with interference from the hydrophilic walls that wet the contact layers) is formed instead of the LDL phase. 
We can see that the influence of attractive walls on the behavior of the system when compared to bulk becomes smaller as 
$L_x$ increases. This can be explained by the lesser relevance of the contribution of a local chemical potential to the global 
chemical potential as the size of the confined system increases. This fact is confirmed by the subtly late filling of the lattice, 
as the chemical potential is high, for higher values of $L_x$.
In figure \ref{fig:17}~(b), the behavior of the average number of hydrogen bonds per site as a 
function of the chemical potential (specifically increasing route of $\overline{\mu}$) is shown for the different 
confinement degrees simulated at $\overline{T}$=~0.30. For all sizes, at high $\overline{\mu}$, in the HDL phase 
(full lattice, $\rho$=~1), the average number of links per site is 2. This is the same value found for the HDL 
phase in the bulk system. 

\begin{figure}[H]
\begin{center}
\subfigure{%
\vspace{1.25cm}
\begin{overpic}[scale=0.95,unit=1mm]{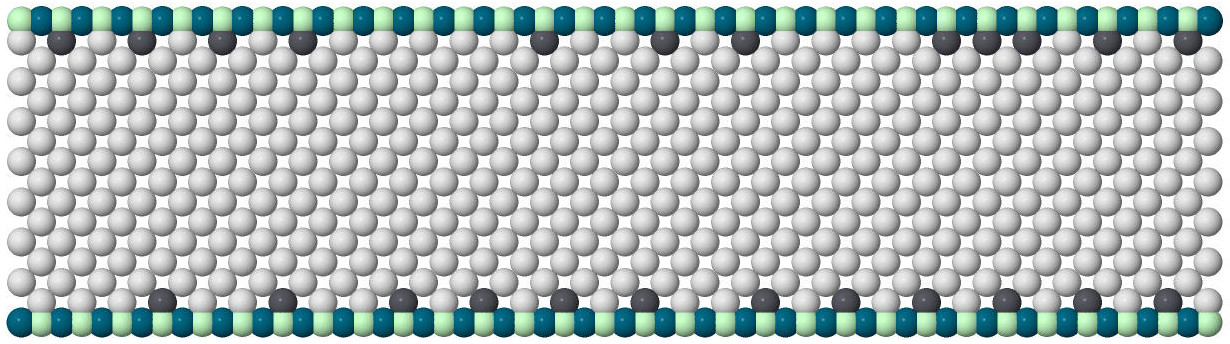}
\put(-4.75,23.0){{\parbox{0.4\linewidth}{
(a)
}}}
\end{overpic}}\hfill
\subfigure{%
\vspace{1.25cm}
\begin{overpic}[scale=0.95,unit=1mm]{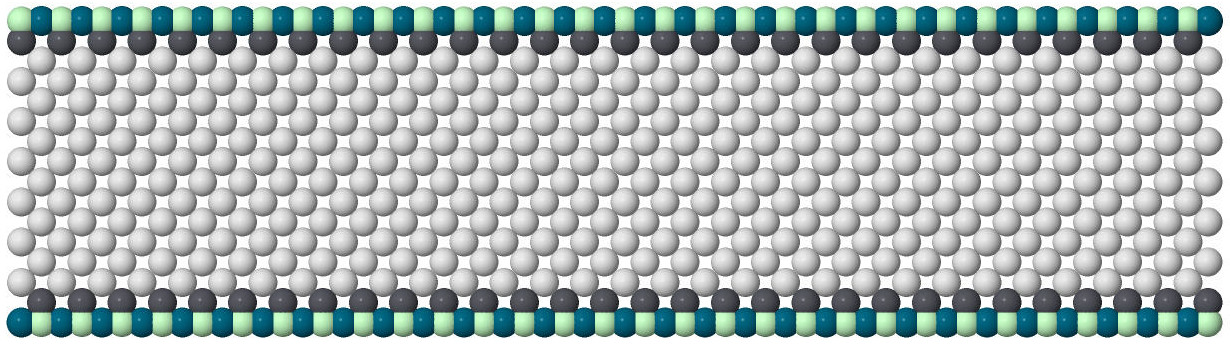}
\put(-4.75,23.0){{\parbox{0.4\linewidth}{
(b)
}}}
\end{overpic}}\hfill
\subfigure{%
\vspace{1.25cm}
\hspace{0.075cm}
\begin{overpic}[scale=1.0,unit=1mm]{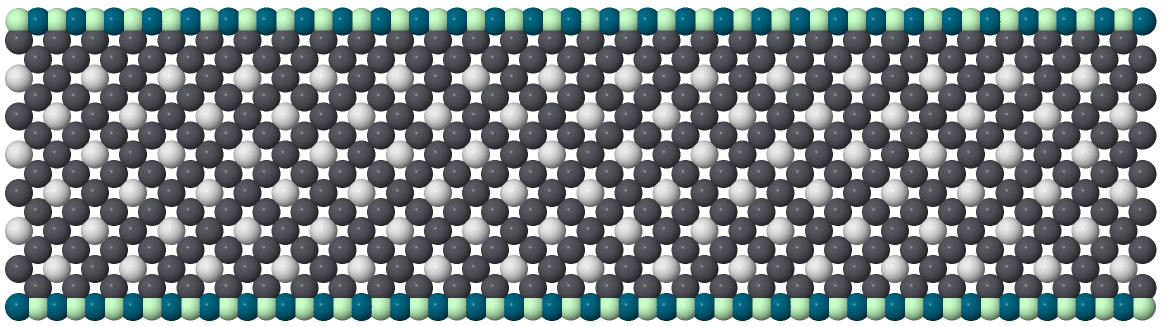}
\put(-4.75,23.0){{\parbox{0.4\linewidth}{
(c)
}}}
\end{overpic}}
\caption{ Photos of the 14x30 confined lattice in the x-y plane at $\overline{T}$=~0.30, showing its phases: in 
(a)~beginning of alternate filling of the lattice ($\overline{\mu}$=~-7.75), (b)~WTT ($\overline{\mu}$=~-4.0) and (c)~ILDL ($\overline{\mu}$=~0.75). }
\label{fig:18}
\end{center}
\end{figure}

\begin{figure}[H]
\begin{center}
\includegraphics[scale=0.225]{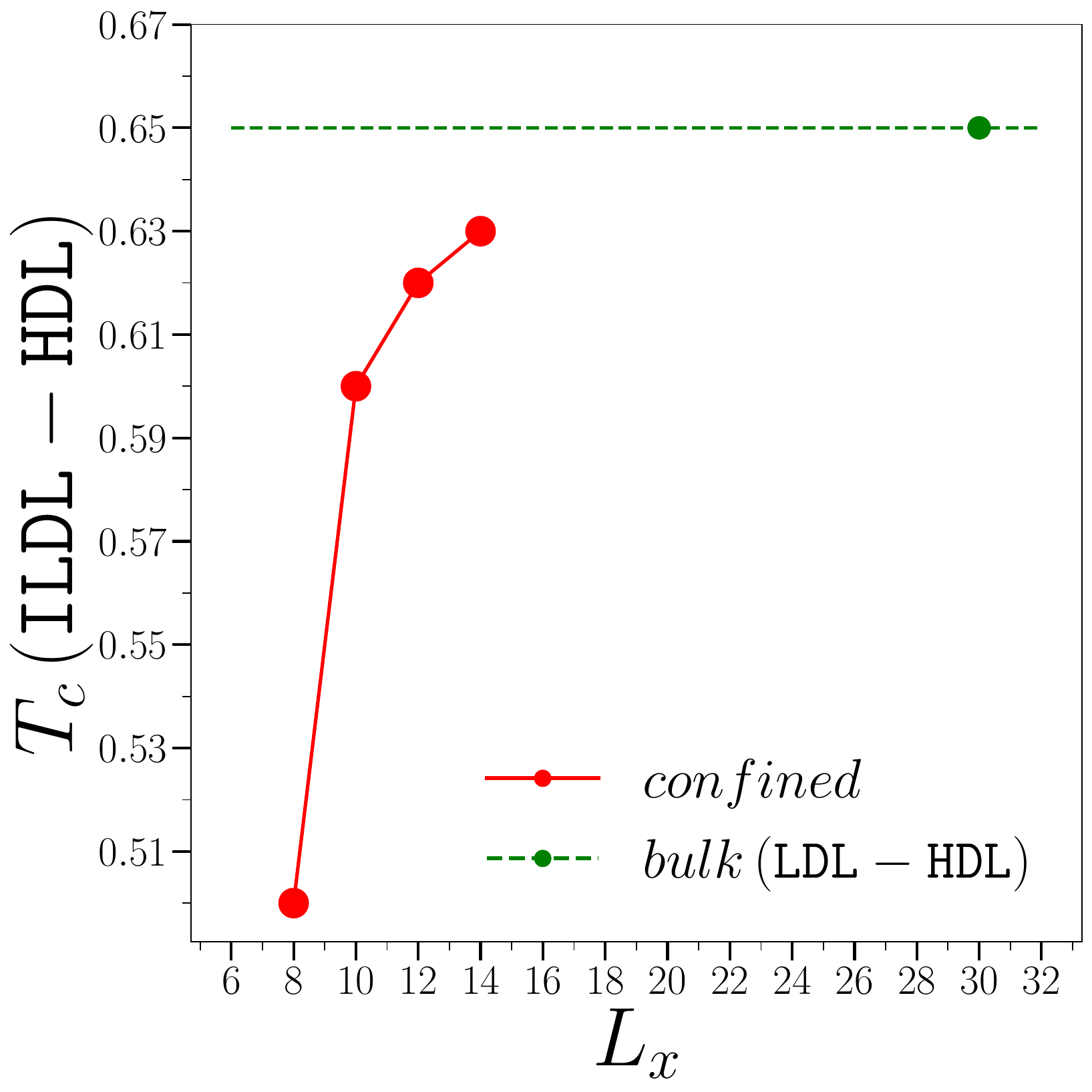}
\caption{ Critical temperatures of the transitions ILDL-HDL, in the confined system, and LDL-HDL, in the bulk system, as a function of $L_x$. }
\label{fig:19}
\end{center}
\end{figure}

As the degree of confinement decreases, the behavior of the bulk system is expected to recover. 
Despite the density of the \textit{wetting} phase presenting a value that is smaller and smaller as the system grows, 
tending to zero in the limit where $L_x$ is large enough, and the decrease in the density interval of the ILDL phase, 
with a tendency to reestablish the bulk LDL phase, where $\rho$=~0.75, the increase of $L_x$ does not resume 
the behavior of the unconfined system. Thus, even for the largest simulated size, $L_x$=~14, the LDL phase, as seen 
in the bulk system, is not recovered, and the WTT phase is still present (figure \ref{fig:18}). 
Thus, we observe the transition between the ILDL and HDL phases, instead of the LDL-HDL transition, with $T_{c\,(ILDL-HDL)} < T_{c\,(LDL-HDL)}$.

Can then say that the behavior of the bulk system can only be recovered for large enough lattices, 
so that the influence of attractive walls, and consequent structural frustration, becomes negligible.

\begin{figure}[H]
\begin{center}
\includegraphics[scale=0.225]{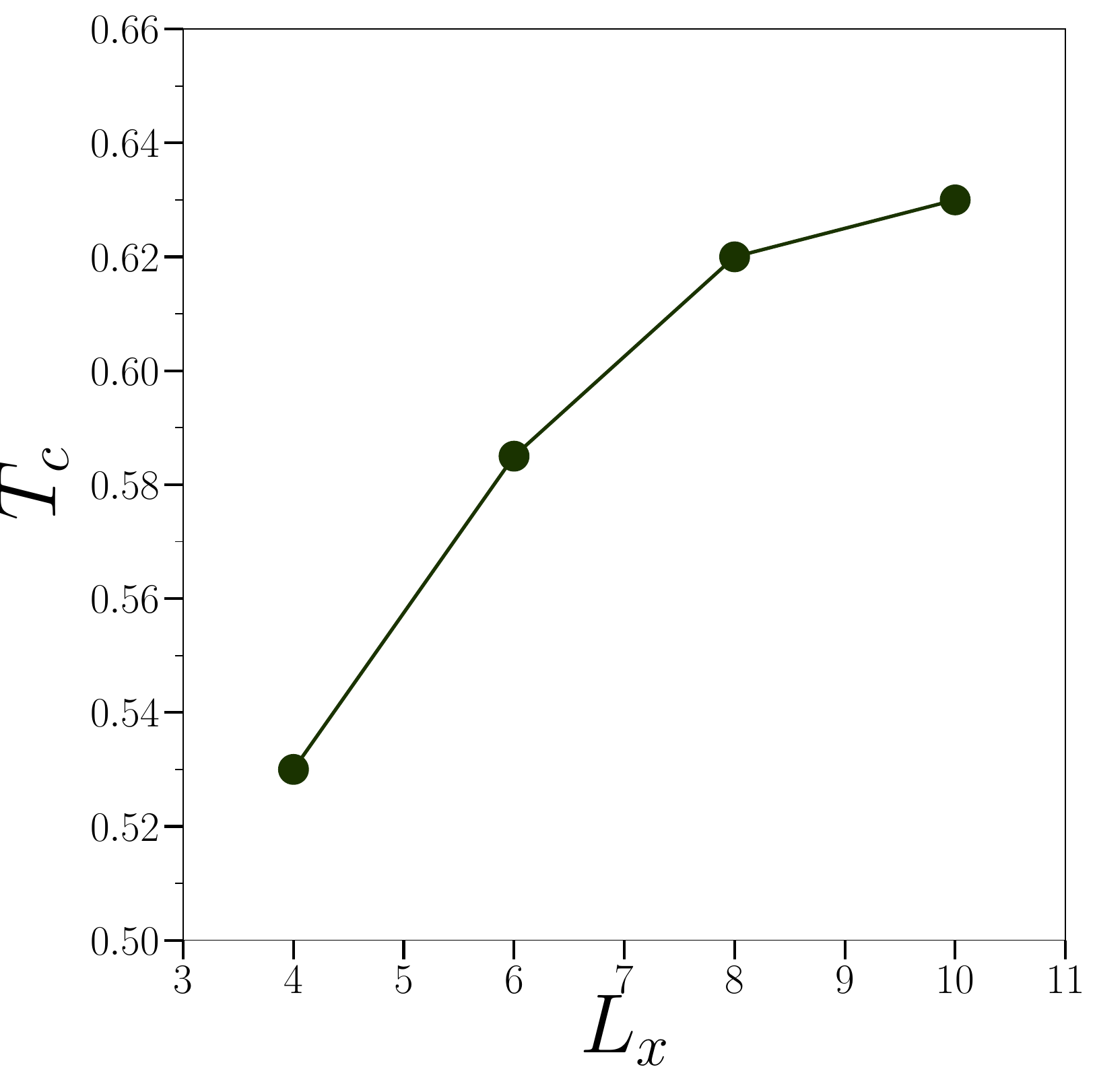}
\caption{ Critical temperatures of the transitions LDL-HDL, in the confined system by hydrophobic walls, and LDL-HDL, 
in the bulk system, as a function of $L_x$ \cite{Fonseca2019}. Reproduced from reference~\cite{Fonseca2019}. }
\label{fig:20}
\end{center}
\end{figure}
The ALGM model for confinement was previously investigated for the case of hydrophobic confinement, 
where wall particles are not allowed to form hydrogen bonds with fluid particles~\cite{Fonseca2019}. 
The hydrophobic confinement, in addition to presenting the phases observed in the unconfined system, 
induces the emergence of a new phase where there is gas in contact with the walls, while a liquid phase 
is established in the central layers, the ``dewetting'' phase, although no drying transition is present. 
Figure \ref{fig:20} brings an important result of this system: the confined fluid exhibits a critical 
liquid-liquid temperature (LDL-HDL) that increases with increasing distance between the confining walls, 
but does not reach the value of bulk, since confining walls affect criticality even for the largest system sizes.

Fluid-wall interaction and geometric constraint are determining factors in trust, as they can affect the 
organization of water molecules. Depending on the structural specificity, confinement can lead to competition 
between the structure imposed by the wall on the fluid and its intrinsic structure. This structural frustration 
can change the orientation and number of hydrogen bonds that water molecules can form. And discrepant behaviors, 
resulting from these frustrated or concurrent interactions, can be observed when comparing the confined system with bulk~\cite{Patricia2018}.

\begin{figure}[H]
\begin{center}
\subfigure {
\begin{overpic}[scale=0.30,unit=1mm]{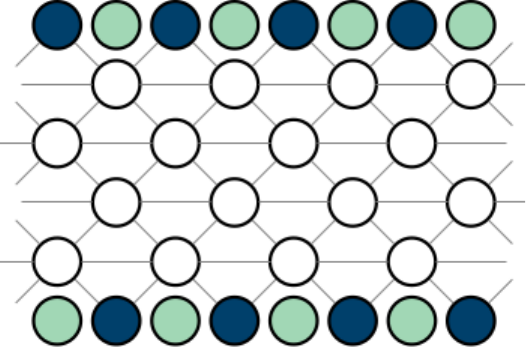}
\put(9.15,-6.0){{\parbox{0.4\linewidth}{
$\overline{\mu}$= -9
}}}
\end{overpic}}\hfill
\subfigure {
\begin{overpic}[scale=0.30,unit=1mm]{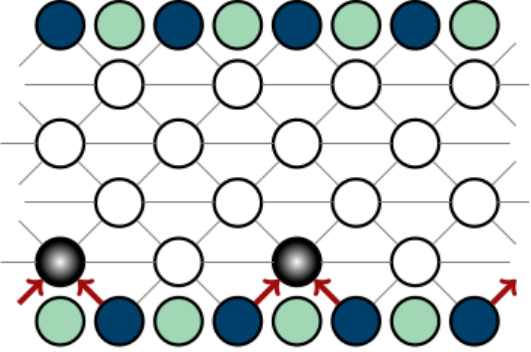}
\put(9.15,-6.0){{\parbox{0.4\linewidth}{
$\overline{\mu}$= -8
}}}
\end{overpic}}\hfill
\subfigure {
\begin{overpic}[scale=0.30,unit=1mm]{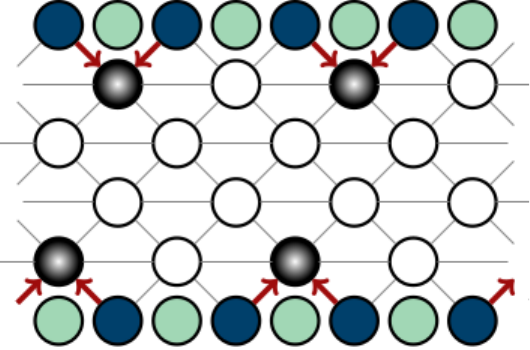}
\put(9.15,-6.0){{\parbox{0.4\linewidth}{
$\overline{\mu}$= -7
}}}
\end{overpic}}\hfill
\vspace{1.0cm}
\subfigure {
\begin{overpic}[scale=0.30,unit=1mm]{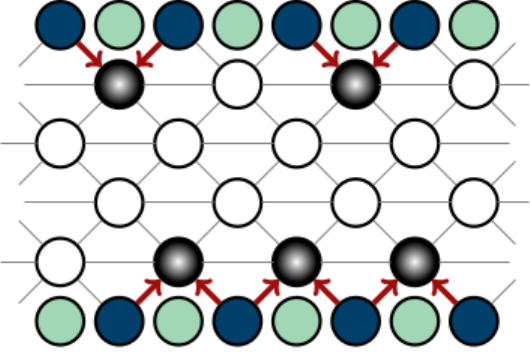}
\put(9.15,-6.0){{\parbox{0.4\linewidth}{
$\overline{\mu}$= -6
}}}
\end{overpic}}\hfill
\subfigure {
\begin{overpic}[scale=0.30,unit=1mm]{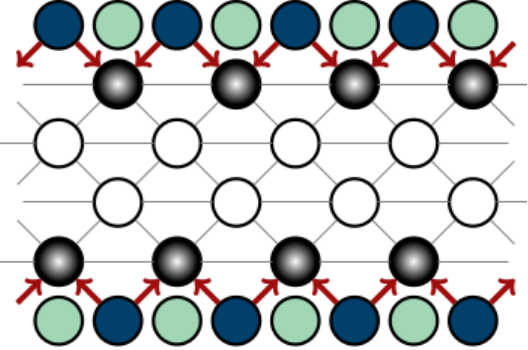}
\put(9.15,-6.0){{\parbox{0.4\linewidth}{
$\overline{\mu}$= -4
}}}
\end{overpic}}\hfill
\subfigure {
\begin{overpic}[scale=0.30,unit=1mm]{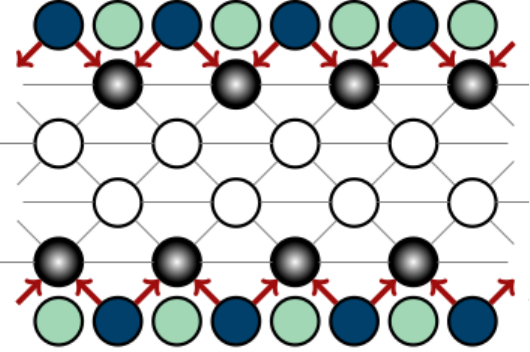}
\put(9.15,-6.0){{\parbox{0.4\linewidth}{
$\overline{\mu}$= -2.75
}}}
\end{overpic}}\hfill
\vspace{1.00cm}
\subfigure {
\begin{overpic}[scale=0.30,unit=1mm]{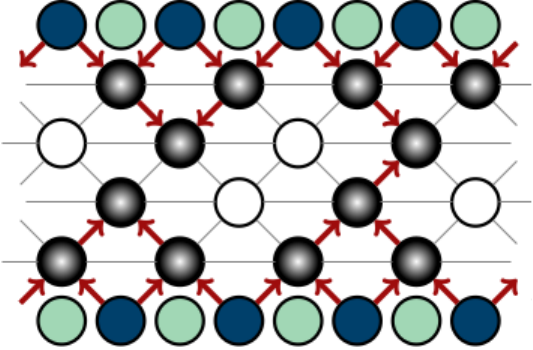}
\put(9.15,-6.0){{\parbox{0.4\linewidth}{
$\overline{\mu}$= -0.75
}}}
\end{overpic}}\hfill
\vspace{0.05cm}
\subfigure {
\begin{overpic}[scale=0.30,unit=1mm]{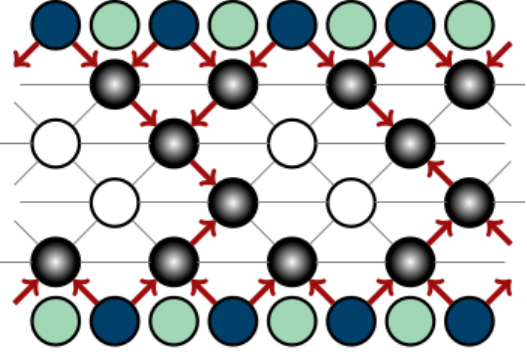}
\put(9.15,-6.0){{\parbox{0.4\linewidth}{
$\overline{\mu}$= 0
}}}
\end{overpic}}\hfill
\subfigure {
\begin{overpic}[scale=0.30,unit=1mm]{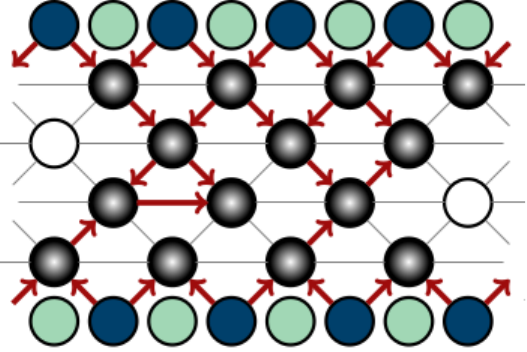}
\put(9.15,-6.0){{\parbox{0.4\linewidth}{
$\overline{\mu}$= +0.75
}}}
\end{overpic}}\hfill
\subfigure {
\begin{overpic}[scale=0.30,unit=1mm]{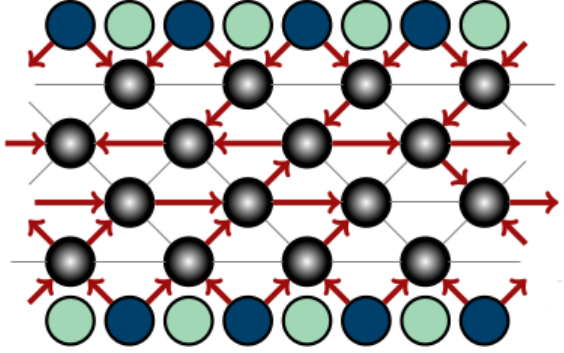}
\put(9.15,-6.0){{\parbox{0.4\linewidth}{
$\overline{\mu}$= 4
}}}
\end{overpic}}\hfill
\subfigure {
\begin{overpic}[scale=0.30,unit=1mm]{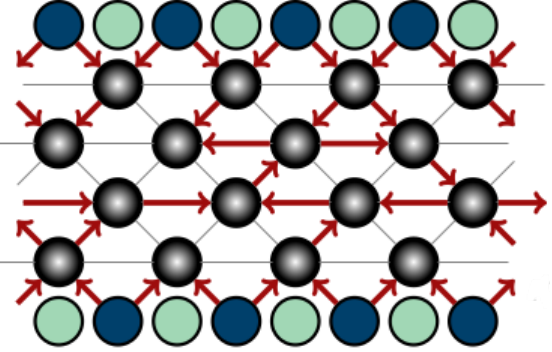}
\put(9.15,-6.0){{\parbox{0.4\linewidth}{
$\overline{\mu}$= 5.25
}}}
\end{overpic}}
\vspace{0.575cm}
\caption{ Illustration of the hydrogen bonds (red arrows) of the 4x4 confined lattice in the x-y plane at 
$\overline{T}$=~0.30 and under increasing chemical potential evolution. }
\label{fig:21}
\end{center}
\end{figure}

As a consequence of frustration in geometry or forces, new structures can be 
observed~\cite{evans2019unified,shaat2019fluidity,wu2004composite}. 
Due to the difficulty of propagating the packing of the confining matrix along the confined fluid, 
there are local structures distinct from the configuration of the system as a whole. Layers in contact 
with the walls may have structural configurations, and also thermodynamics, different from the inner layers, 
resulting in a non-homogeneous structural behavior 
(with hydrogen bonds that vary with the distance from the confining surface)~\cite{rieth2019hydrogen,grason2016perspective}.

Often, as a consequence of nanoscale confinement, water molecules at protein interfaces are stymied of 
hydrogen bonding opportunities. This impossibility can lead to functional differences and give rise to 
considerable biological effects \cite{ferreiro2018frustration}.

\begin{figure}[H]
\begin{center}
\subfigure {
\begin{overpic}[scale=0.30,unit=1mm]{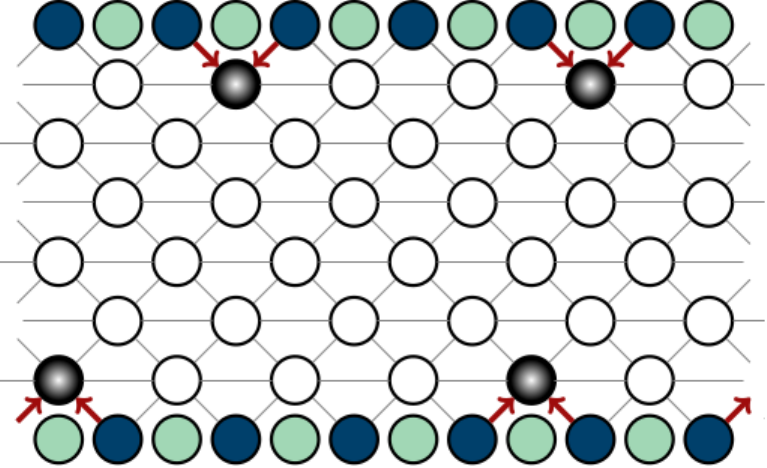}
\put(15.5,-6.0){{\parbox{0.4\linewidth}{
$\overline{\mu}$= -8
}}}
\end{overpic}}\hfill
\subfigure {
\begin{overpic}[scale=0.30,unit=1mm]{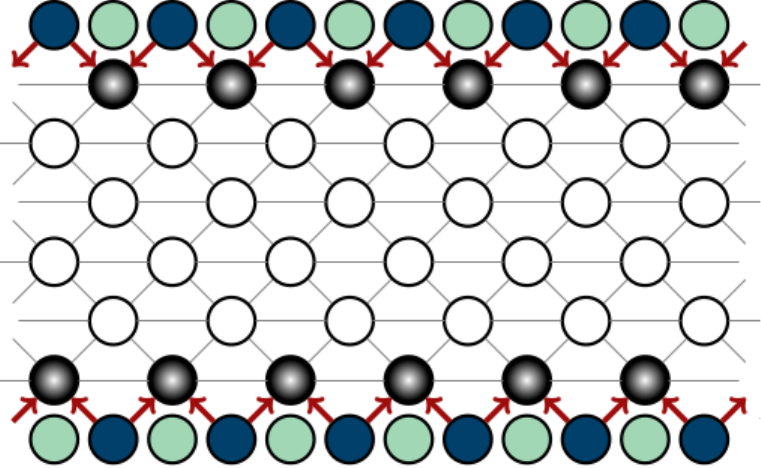}
\put(15.5,-6.0){{\parbox{0.4\linewidth}{
$\overline{\mu}$= -4
}}}
\end{overpic}}\hfill
\vspace{1.0cm}
\subfigure {
\begin{overpic}[scale=0.30,unit=1mm]{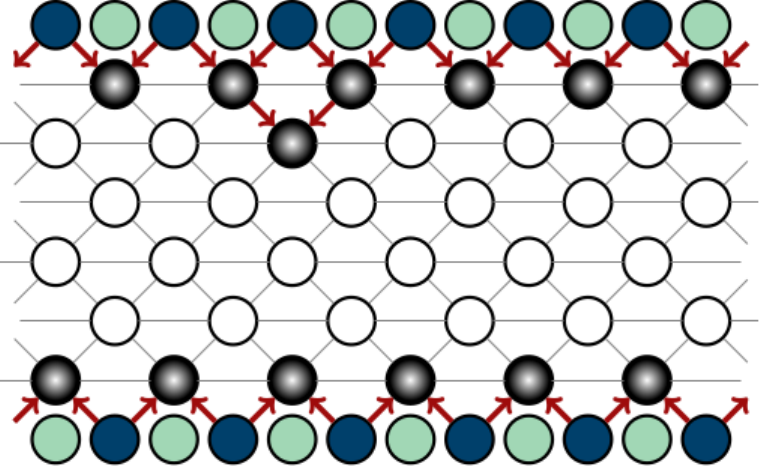}
\put(15.5,-6.0){{\parbox{0.4\linewidth}{
$\overline{\mu}$= -2.75
}}}
\end{overpic}}\hfill
\subfigure {
\begin{overpic}[scale=0.30,unit=1mm]{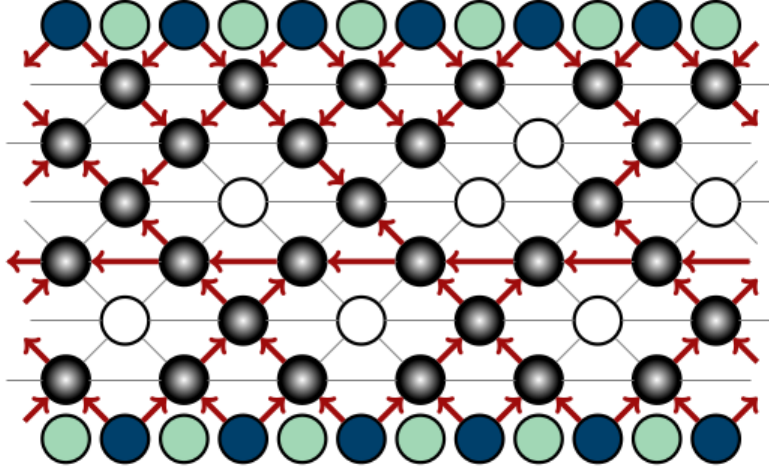}
\put(15.5,-6.0){{\parbox{0.4\linewidth}{
$\overline{\mu}$= 0
}}}
\end{overpic}}\hfill
\vspace{1.0cm}
\subfigure {
\begin{overpic}[scale=0.30,unit=1mm]{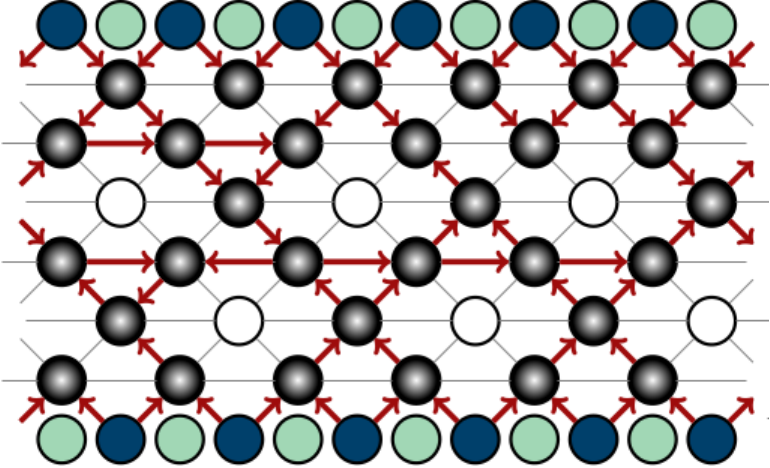}
\put(15.5,-6.0){{\parbox{0.4\linewidth}{
$\overline{\mu}$= 0.75
}}}
\end{overpic}}\hfill
\subfigure {
\begin{overpic}[scale=0.30,unit=1mm]{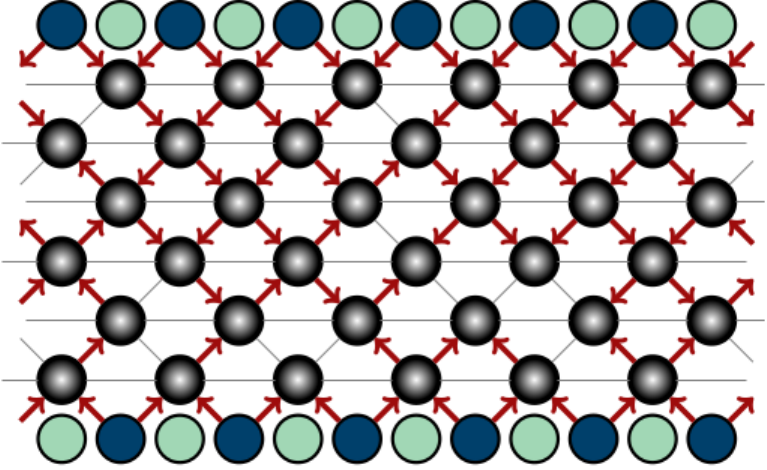}
\put(15.5,-6.0){{\parbox{0.4\linewidth}{
$\overline{\mu}$= 4
}}}
\end{overpic}}
\vspace{-0.375cm}
\caption{ Illustration of the hydrogen bonds (red arrows) of the 6x6 confined lattice 
in the x-y plane at $\overline{T}$=~0.30 and under increasing chemical potential evolution. }
\label{fig:22}
\end{center}
\end{figure}

Although these are finite systems, in figures \ref{fig:21} and \ref{fig:22}, 
the simulated confinements 4x4 and 6x6 help us to elucidate the behavior of the confined water of the our hydrophilic 
confinement model. The attractive wall competes with water-water interactions and are energetically more favorable in 
forming hydrogen bonds with water molecules (fluid-wall interaction is more attractive than fluid-fluid interaction). 
We observed how the attempt at organization imposed by the confining matrix has its impact diminished as the distance 
from the surface increases. Therefore, the propagation of the structure imposed by the wall along the inner layers becomes smaller.

In the 4x4 lattice, with the increase in the chemical potential and consequent increase in the density of the system, 
we observe how this competition affects the organization of hydrogen bonds in the central layers. We noticed that due 
to the extreme confinement of attractive walls, the water molecules, without enough space to rearrange themselves, are 
configured in a structure where the hydrogen bonds have very different orientations from those observed in the LDL and HDL 
phases of the bulk system (see figure \ref{fig:6}). 
In contrast, in the 6x6 lattice, which has a lower degree of confinement (greater space for water molecules to reorient themselves), 
the frustration displayed is notably less, and the configurations of the hydrogen bonds resemble the configurations 
of the phases of the unconfined system.

\section{Conclusion}

Using a two-dimensional associative gas lattice model for confinement, we investigated the behavior of the phases 
of a water-like fluid under hydrophilic confinement. Hydrophilic confinement allows the confining matrix to form 
hydrogen bonds with water molecules. We observed the presence of the GAS and HDL phases in the unconfined system, 
in addition to new structures and the early filling of the lattice due to the presence of attractive walls.

The attractive wall acts as a higher local chemical potential, contributing to an increase in the global chemical 
potential of the system. Thus, the gas phase becomes shorter and HDL phase appears at lower chemical potentials, 
leading to a greater occurrence of this phase throughout the variation of the chemical potential. In this way, the 
lattice is filled with water molecules at significantly lower chemical potentials when compared to the bulk system.

A new structure occurs at low chemical potentials and non-extreme confinements, ${L_x}\geq$~2, the wetting layer that gradually 
wets the wall. Unlike the wetting transition, this layer is unique and operates as a chemical potential that promotes water ordering. 
From the wetting layer, a configuration similar to the low-density liquid is formed, but which is interfered with by the attractive walls, 
the ILDL, interfered LDL. Even for the largest simulated size, the LDL phase, as seen in the bulk system, is not recovered, 
and the WTT phase is still present. Therefore, the ILDL-HDL transition was observed, instead of the LDL-HDL transition, 
with $T_c$(ILDL-HDL) <  $T_c$(LDL-HDL). This shift in the transition temperature between the liquid phases is related to the 
same effect observed in the hydrophobic system~\cite{Fonseca2019}, the presence of the confining walls and consequent structural frustration.

Therefore, the investigated confined systems can only recover the behavior of the bulk system for a lattice large enough, 
where the influence of the walls and consequent structural frustration becomes negligible.


\begin{thebibliography}{99}
%
\bibitem{barroug2004interactions} Barroug, A., Kuhn, L., Gerstenfeld, L. and Glimcher, M., Journal Of Orthopaedic Research, 22, 703-708 (2004).

\bibitem{pai2006pharmaceutical} Pai, P., Nair, K., Jamade, S., Shah, R., Ekshinge, V. and Jadhav, N., Current Pharma Esearch Journal, 1, 11-15 (2006).

\bibitem{pantarotto2003immunization} Pantarotto, D., Partidos, C., Hoebeke, J., Brown, F., Kramer, E., Briand, J., Muller, S., Prato, M. and Bianco, A. , Chemistry \& Biology, 10, 961-966 (2003).

\bibitem{kam2005carbon} Kam, N., O Connell, M., Wisdom, J. and Dai, H., Proceedings Of The National Academy Of Sciences, 102, 11600-11605 (2005).

\bibitem{ding2001recent} Ding, R., Lu, G., Yan, Z. and Wilson, M., Journal Of Nanoscience And Nanotechnology, 1, 7-29 (2001).

\bibitem{wang2010natural} Wang, S. and Peng, Y., Chemical Engineering Journal, 156, 11-24 (2010).

\bibitem{babel2003low} Babel, S. and Kurniawan, T., Journal Of Hazardous Materials, 97, 219-243 (2003).

\bibitem{fonseca2017freezing} Fonseca, T., Ternes, P., Oliveira, A. and Barbosa, M., Physicae Organum, 3, 128-138 (2017).

\bibitem{Driemeier2012} Driemeier, C., Mendes, F. and Oliveira, M., Cellulose, 19, 1051-1063 (2012).

\bibitem{Gelb2006} Gelb, L., Gubbins, K., Radhakrishnan, R. and Sliwinska-Bartkowiak, M., Reports On Progress In Physics, 62, 1573 (2006).

\bibitem{Deville2006} Deville, S., Saiz, E., Nalla, R. and Tomsia, A., Science, 311, 515-518 (2006).

\bibitem{ZheWang2015} Wang, Z., Ito, K., Leão, J., Harriger, L., Liu, Y. and Chen, S., The Journal Of Physical Chemistry Letters, 6, 2009-2014 (2015).

\bibitem{Gavazzoni2017} Gavazzoni, C., Giovambattista, N., Netz, P. and Barbosa, M., The Journal Of Chemical Physics, 146, 234509 (2017).

\bibitem{Zangi2003} Zangi, R. and Mark, A., The Journal Of Chemical Physics, 119, 1694-1700 (2003).

\bibitem{Giovambattista2009_01} Giovambattista, N., Rossky, P. and Debenedetti, P., Phys. Rev. Lett., 102, 050603 (2009,2).

\bibitem{Nanok2009} Nanok, T., Artrith, N., Pantu, P., Bopp, P. and Limtrakul, J., The Journal Of Physical Chemistry A, 113, 2103-2108 (2009).

\bibitem{Krott2014} Krott, L. and Barbosa, M., Phys. Rev. E, 89, 012110 (2014).

\bibitem{Bordin2014} Bordin, J., Krott, L. and Barbosa, M., The Journal Of Physical Chemistry C, 118, 9497-9506 (2014).

\bibitem{Findenegg2008} Findenegg, G., Jahnert, S., Akcakayiran, D. and Schreiber, A., Chem Phys Chem, 9, 2651-2659 (2008).

\bibitem{Allenhnert2008} James T Allenhnert, S., Chavez, F., Schaumann, G., Schreiber, A., SJaonhoff, M. and Findenegg, G., Physical Chemistry Chemical Physics, 10, 6039-6051 (2008).

\bibitem{Erko2011} Erko, M., Findenegg, G., Cade, N., Michette, A. and Paris, O., Physical Review B, 84, 104205 (2011).

\bibitem{Morishige1999} Morishige, K. and Kawano, K., The Journal Of Chemical Physics, 110, 4867-4872 (1999).

\bibitem{Gallo2007} Gallo, P. and Rovere, M., Physical Review E, 76, 061202 (2007).

\bibitem{Pizio2009} Pizio, O., Dominguez, H., Pusztai, L. and Sokolowski, S., Physica A: Statistical Mechanics And Its Applications, 388, 2278-2288 (2009).

\bibitem{Koga2001} Koga, K., Gao, G., Tanaka, H. and Zeng, X., Nature, 412, 802-805 (2001).

\bibitem{Hummer2001} Hummer, G., Rasaiah, J. and Noworyta, J., Nature, 414, 188-190 (2001).

\bibitem{Bellissent1996} Bellissent-Funel, M., Sridi-Dorbez, R. and Bosio, L., The Journal Of Chemical Physics, 104, 10023-10029 (1996).

\bibitem{Zanotti2005} Zanotti, J., Bellissent-Funel, M. and Chen, S., EPL (Europhysics Letters, 71, 91 (2005).

\bibitem{Koga2005} Koga, K. and Tanaka, H., The Journal Of Chemical Physics, 122, 104711 (2005).

\bibitem{Choudhury2010} Choudhury, N., The Journal Of Chemical Physics, 132, 064505 (2010).

\bibitem{Meyer1999} Meyer, M. and Stanley, H., The Journal Of Physical Chemistry B, 103, 9728-9730 (1999).

\bibitem{Kumar2005_01} Kumar, P., Buldyrev, S., Sciortino, F., Zaccarelli, E. and Stanley, H., Physical Review E, 72, 021501 (2005).

\bibitem{Molinero2012} Moore, E., Allen, J. and Molinero, V., The Journal Of Physical Chemistry C, 116, 7507-7514 (2012).

\bibitem{Deschamps2010} Deschamps, J., Audonnet, F., Brodie-Linder, N., Schoeffel, M. and Alba-Simionesco, C., Physical Chemistry Chemical Physics, 12, 1440-1443 (2010).

\bibitem{Stapf1995} Stapf, S. and Kimmich, R., The Journal Of Chemical Physics, 103, 2247-2250 (1995,8).

\bibitem{STEYTLER1983} STEYTLER, J., J. Phys. Chem., 87, 2458-2459 (1983).

\bibitem{Liu2006} Liu, E., Dore, J., Webber, J., Khushalani, D., Jahnert, S., Findenegg, G. and Hansen, T., J. Phys.: Cond. Matter, 18, 10009 (2006).

\bibitem{OVERLOOP1993} Overloop, K. and Vangerven, L., Journal Of Magnetic Resonance, Series A, 101, 179 - 187 (1993).

\bibitem{Morishige2003} Morishige, K. and Iwasaki, H., Langmuir, 19, 2808-2811 (2003).

\bibitem{Baker1997} Baker, J., Dore, J. and Behrens, P., The Journal Of Physical Chemistry B, 101, 6226-6229 (1997).

\bibitem{Chen2004} Faraone, A., Liu, L., Mou, C., Yen, C. and Chen, S., The Journal Of Chemical Physics, 121, 10843-10846 (2004).

\bibitem{Xu2005} Xu, L., Kumar, P., Buldyrev, S., Chen, S., Poole, P., Sciortino, F. and Stanley, H., Proc Natl Acad Sci USA, 102, 16558-16562 (2005,11).

\bibitem{Jelassi2010} Jelassi, J., Castricum, H., Bellissent-Funel, M., Dore, J., Webber, J. and Sridi-Dorbez, R., Phys. Chem. Chem. Phys., 12, 2838 (2010).

\bibitem{Patricia2018} P. Ternes, E. Salcedo and M. C. Barbosa, Phys. Rev. E 97, 033104 (2018).

\bibitem{Beckstein2003} O. Beckstein and M. S. P. Sansom, Proc. Natl. Acad. Sci. U.S.A. 100, 7063 (2003).

\bibitem{Tajkhorshid2002} E. Tajkhorshid, P. Nollert, M. O. Jensen, L. J. W. Miercke, J. O Connel, R. M. Stroud, and K. Schulten, Science 296, 525 (2002).

\bibitem{Murata2000} K. Murata, K. Mitsuoka, T. Hirai, T. Walz, P. Agre, J. B. Heymann, A. Engel, and Y. Fujiyoshi, Nature 407, 599 (2000).

\bibitem{Fu2000} D. Fu, A. Libson, L. J. W. Miercke, C. Weitzman, P. Nollert, J. Krucinski, and R. M. Stroud, Science 290, 481 (2000).

\bibitem{Sui2001} H. Sui, B.-G. Han, J. K. Lee, P. Wallan, and B.  K. Jap, Nature 414, 872 (2001).

\bibitem{Harpham2004} M. R. Harpham, B. M. Ladanyi, N. E. Levinger, and K. W. Herwig, J. Chem. Phys. 121, 7855 (2004).

\bibitem{Keliu2017} K. Wu, Zhangxin Chen, J. Li, X. Li, J. Xu, and X. Dong, PNAS 114, 3358 (2017).

\bibitem{Saraswat2018} V. Saraswat, R. M. Jacobberger, J. S. Ostrander, C. L. Hummell, A. J. Way,  J. Wang,  M. T. Zanni, and M. S. Arnold, ACS Nano 12, 7855 (2018).

\bibitem{Holt2006} J. K. Holt, H. G. Park, Y. Wang, M. Stadermann, A. B. Artyukhin, C. P. Grigoropoulos,  A. Noy,  and O. Bakajin, Science 312, 1034 (2006). 

\bibitem{Kohler2019} M. H. Kohler, J. R. Bordin and M. C. Barbosa, Molecular Liquids (2019).

\bibitem{Li2005} Li Liu, Sow-Hsin Chen, Antonio Faraone, Chun-Wan Yen, and Chung-Yuan Mou, Phys. Rev. Lett. 95, 117802 (2005).

\bibitem{Xu2005} Limei Xu, Pradeep Kumar, S. V. Buldyrev, S.-H. Chen, P. H. Poole, F. Sciortino, and H. E. Stanley, PNAS 102, 16558 (2005).

\bibitem{Szortyka2010134904} Szortyka, M., Girardi, M., Henriques, V. and Barbosa, M., The Journal Of Chemical Physic, 132, 134904 (2010)

\bibitem{Gallo2010} P. Gallo, M. Rovere and S.-H. Chen, J. Phys. Chem. Lett. 1, 729 (2010).

\bibitem{Cervery2016}S. Cervery,  F.  Mallamace, J. Swenson, M.  Vogel, and L.  Xu, Chem. Rev. 116, pp 7608 (2016).

\bibitem{Stillinger1973}F. H. Stillinger, J. Solution Chem. 2, 141 (1973).

\bibitem{Lum1999} K. Lum, D. Chandler and J. D. Weeks, Int. J. Food Prop. 103, 4570 (1999).

\bibitem{Jensen2003} T. R. Jensen, M. O. Jensen, N. Reitzel, K. Balashev, G. H. Peters, K. Kjaer, 1 and T. Bjornholm, Phys. Rev. Lett. 90, 086101 (2003).

\bibitem{Schwendel2003} D. Schwendel, T. Hayashi, R. Dahint, A. Pertsin, ~ M. Grunze, R. Steitz, and F. Schreibe, Langmuir 19, 2284 (2003).

\bibitem{Zhang2018} Z. Zhang, S. Ryu, Y. Ahn and Joonkyung Jang, Phys. Chem. Chem. Phys. 20, 30492 (2018).

\bibitem{Wallqvistt1995} A. Wallqvistt and B. J. Berne, J. Phys. Chem. 1995, 99, 2893 (1995).

\bibitem{Huang2003} X. Huang, C. J. Margulis, B. J. Berne, Proc. Natl. Acad. Sci. 100, 11953 (2003). 

\bibitem{Giovambattista2006} N. Giovambattista, P. J. Rossky, and P. G. Debenedetti, Phys. Rev. E 73, 41604 (2006).

\bibitem{Giovambattista2009} N. Giovambattista, P. J. Rossky, and P. G. Debenedetti, Phys. Chem. B 113, 13723 (2009).

\bibitem{ten2002} P. R. ten Wolde, and D.Chandler, D., Proc. Natl. Acad. Sci.  99, 6539 (2002).

\bibitem{Liu2005} P. Liu, X. Huang, R.  Zhou, and B. J. Berne, Nature  437, 159 (2005). 

\bibitem{Hua2007} L. Hua, X. Huang, P. Liu, R. Zhou and B. J. Berne, J. Phys. Chem. B 111, 9069 (2007).

\bibitem{Zhou2004} R.  Zhou, X. Huang, C. J. Margulis and B. J. Berne, Science 305, 1605 (2004).

\bibitem{MacCullum2007} J. L. MacCullum, M. S. Maghaddam, H. S. Chan, D. P. Tieleman, Proc. Natl. Acad. Sci. 104, 6206 (2007).

\bibitem{Cicero2008} G. Cicero, J. C. Grossman, E. Schwegler, F. Gygi and G. Galli, J. Am. Chem. Soc. 130, 1871 (2008).

\bibitem{Scodinu2002} A. Scodinu, and J T. Fourkas, J. Chem. Phys. B  106, 10292 (2002).

\bibitem{Kim2003} H. I. Kim, J. G. Kushmmerick, J. E. Houston, B. C. Bunker, Langmuir 19, 9271 (2003).

\bibitem{Bagchi2005} B. Bagchi, Chem. Rev. 105, 3197 (2005).

\bibitem{Henriques2005} V. B. Henriques and M. C. Barbosa, Phys. Rev. E. 71, 031504 (2005).

\bibitem{Szortyka2007} M. M. Szortyka and M. C. Barbosa, Physica A 380, 27 (2007).

\bibitem{Szortyka2009} M. M. Szortyka, V. B. Henriques, M. Girardi and M. C. Barbosa, J. Chem. Phys. 130, 184902 (2009).

\bibitem{Kumar2005} P. Kumar, S. V. Buldyrev,  F. W. Starr,  N. Giovambattista  and H. E.  Stanley, Phys. Rev. E 72, 051503 (2005).

\bibitem{Moore2012} E. B. Moore, J. T. Allen, and V. Molinero, J. Phys. Chem. C 116, 7507 (2012).

\bibitem{Farimani2016} A. B.  Farimani and N. R. Aluru, J. Phys. Chem. C  120, 23763 (2016).

\bibitem{Li2006} X. Li, J. Li,  M.  Eleftheriou  and R.  Zhou, J. AM. CHEM. SOC. J. Am. Chem. Soc.  128, 12439 (2006).

\bibitem{Bordin2014} J. R. Bordin, L. B. Krott and M. C. Barbosa, J. Phys. Chem. C 118, 9497 (2014).

\bibitem{Leoni2014} F. Leoni, and G. Franzese, J. Chem. Phys. 141, 174501 (2014).

\bibitem{Pestana2018} L. R. Pestana,  L.  E. Felberg,  and T. Head-Gordon, ACS Nano  12, 448 (2018).

\bibitem{Mendonca2019} B. H. S. Mendonça, D. N. de Freitas, M. H. Kohler, M. C. Barbosa and A. B. de Oliveira, Physica A 517, 491 (2019). 

\bibitem{Fonseca2019} T. O. Fonseca, M. M. Szortyka, P. Ternes, C. Gavazzoni, A. B. de Oliveira and M. C. Barbosa, Science China Physics, Mechanics \& Astronomy 62, 10 (2019).

\bibitem{evans2019unified} Evans, R., Stewart, M. and Wilding, N., Proceedings Of The National Academy Of Sciences, 116, 23901-23908 (2019).

\bibitem{shaat2019fluidity} Shaat, M. and Zheng, Y., Scientific Reports, 9, 1-12 (2019).

\bibitem{verdaguer2006molecular} Verdaguer, A., Sacha, G., Bluhm, H. and Salmeron, M., Chemical Reviews, 106, 1478-1510 (2006).

\bibitem{malani2009influence} Malani, A., Ayappa, K. and Murad, S., The Journal Of Physical Chemistry B, 113, 13825-13839 (2009).

\bibitem{wu2004composite} Wu, Y., Cheng, G., Katsov, K., Sides, S., Wang, J., Tang, J., Fredrickson, G., Moskovits, M. and Stucky, G., Nature Materials, 3, 816-822 (2004).

\bibitem{rieth2019hydrogen} Rieth, A., Hunter, K., Dinca, M. and Paesani, F., Nature Communications, 10, 1-7 (2019).

\bibitem{grason2016perspective} Grason, G. M., The Journal Of Chemical Physics, 145, 110901 (2016).

\bibitem{ferreiro2018frustration} Ferreiro, D. U., Komives, E. A. and Wolynes, P. G., Current Opinion In Structural Biology, 48, 68-73 (2018).


%
\end{thebibliography}
\end{document}